\documentclass[letterpaper,twocolumn,10pt]{article}
\usepackage{usenix}

\usepackage{url}
\makeatletter
\g@addto@macro\UrlBreaks{\do\/\do\-}
\makeatother
\usepackage{tikz}
\usepackage{amssymb}
\usepackage{amsmath}

\usepackage{tabularx}
\usepackage{array}       
\usepackage{booktabs}    
\usepackage{comment}
\usepackage{graphicx}
\usepackage{float}
\usepackage{subcaption}
\usepackage{balance}

\usepackage[dvipsnames]{xcolor}

\usepackage{xspace, xparse}

\NewDocumentCommand\am{g}{%
  \IfNoValueF{#1}{{\color{blue} \textbf{(AM: #1)}}}%
  \IfNoValueT{#1}{{\color{blue} \textbf{(AM)}}}%
}

\NewDocumentCommand\mm{g}{%
  \IfNoValueF{#1}{{\color{blue} \textbf{(MM: #1)}}}%
  \IfNoValueT{#1}{{\color{blue} \textbf{(MM)}}}%
}

\newcommand{\subhead}[1]{\vspace{1pt}\noindent{\textbf{#1.}}}

\newcommand{\bm}[1]{\mathbf{#1}} 

\newcommand\raiseT[2]{%
\setbox0\hbox{$#1{#2}$}\raise\dp0\box0}

\usepackage{amsthm}
\newtheorem{theorem}{Theorem}

\begin{document}

\date{}

\title{\Large \bf On Improving Robustness of Deepfake  Image Detectors}

 \author{
 {\rm Abu Taib Mohammed Shahjahan}\\
 Concordia University, Montreal, Canada
 \and
 {\rm Mohammad Mannan}\\
 Concordia University, Montreal, Canada
 \and
 {\rm Abdessamad Ben Hamza}\\
 Concordia University, Montreal, Canada
 \and
 {\rm Amr Youssef}\\
 Concordia University, Montreal, Canada
 }
\pagestyle{empty}

\maketitle

\begin{abstract}
The rapid advancement of Generative AI has introduced remarkable opportunities while simultaneously raising critical concerns regarding content authenticity. While recent work has increasingly focused on improving the generalization of deepfake detectors across unseen generative models, their robustness against adversarial attacks remains limited. In particular, Abdullah et al.\ (IEEE SP 2024) evaluated eight detectors and demonstrated that most of them exhibit significant performance degradation under adversarial attacks. We also observed the same phenomenon by testing seven most recent state-of-the-art  detectors.  
To address this problem, we propose a unified framework that integrates three complementary design principles without relying on adversarial training data: (i) higher-order statistical modeling in the frequency domain via Discrete Cosine Transform (DCT)-based moment pooling up to fourth order, (ii) content-agnostic feature representations derived from noise residuals, and (iii) cross-scene generalization enforced through patch-level semantic disruption. A key insight underpinning our approach is that adversarial attacks primarily operate on low-order statistics and visual semantics, leaving higher-order residual-frequency characteristics, particularly kurtosis, largely unconstrained. Extensive experiments demonstrate that our method consistently improves robustness across six architecturally diverse detectors. Notably, we achieve up to 88.9\% reduction in recall degradation on current adversarial benchmarks, and improve the best-performing recent detector (Yang et al., IEEE CVPR 2025) from 81.9\% to 97.15\% accuracy under attack. Overall, our method provides a principled, architecture-agnostic approach for improving deepfake detection robustness against current attacks.

\end{abstract}

\section{Introduction}

\noindent The development of generative artificial intelligence (GenAI) has dramatically accelerated the synthesis of media content, such as images and videos. Within GenAI, a specific area of focus lies in the generation of fully synthetic images, commonly referred to as \emph{deepfakes}. Unlike manual creation using graphics applications, well-trained deepfake  generation models excel in both performance and speed of generation. Deepfake  generation by Generative Adversarial Networks (GANs) and Diffusion Models (DMs)~\cite{sohl2015Nonequilibriumthermodynamics, Podell2024sdxl, rombach2022ldm, Liu2023rectifiedflow, Esser2024rectifiedflow, zhang2025diffusion4k} have attracted much public attention due to their visually striking outputs, which are nearly indistinguishable from genuine photographs, even to human observers~\cite{Nightingale2022AISynthesized}. 

Such progress, while remarkable, also introduces significant societal risks, as these synthetic images can be exploited for malicious purposes, including blackmailing, harassment, and large-scale disinformation campaigns. Diffusion model (DM)-based text-to-image frameworks, in particular, heighten these concerns by enabling adversaries to generate highly persuasive synthetic imagery aligned with specific narratives while requiring minimal technical expertise. As the realism and accessibility of these technologies continue to improve, recent studies have warned that deepfakes can erode public trust in digital media, amplify misinformation during politically sensitive events, and inflict significant psychological and societal harm~\cite{alanazi2025unmasking}. 
The rapid proliferation of deepfake generation tools has also raised important legal and ethical concerns related to privacy violations, identity misuse, and the urgent need for effective regulatory frameworks to mitigate abuse~\cite{meskys2020regulating}. These concerns are no longer theoretical, as harmful misuse of generative AI images has already caused severe real-world consequences, including suicide linked to AI-generated morphed nude images and videos~\cite{toi_ai_blackmail_suicide_2025}, acute mental health impacts on young people~\cite{guardian_deepfake_porn_schools_2025}, and political misinformation following the capture of Venezuelan President Maduro~\cite{guardian2026maduroai}.

Due to increasing awareness of such threats and rapid advancements in generative techniques, several tools exist for fully synthetic image detection~\cite{Wang2023dire,Tran2025diffcor,Chu2025fire,Karageorgiou2025anyresolution,Ricker2024detection,Tan2024upsampling,yang2025d3,Cheng2025cospy,Huang2025sida,Wang2023dynamicgraphlearning,Brokman2025manifold,Zheng2024semanticartifacts,Corvi2023intriguing,Wolter2022wavelet,Sarkar2024shadows,Lorch2024landscape}. 
Common approaches include detecting semantic inconsistencies in images, or capturing low-level artifacts introduced by the generative models for detection. For example, geometric inconsistencies in scenes as well as in shadows, perspective  and lighting in an image are used for detection~\cite{Sarkar2024shadows}; low-level artifacts from generative models are captured in the spatial domain~\cite{Wang2024cnngenerated}, or in the spectral domain~\cite{Li2024masksim}. Such detectors, as reported, often achieve very high accuracy in deepfake detection; e.g., $100\%$ accuracy of FIRE~\cite{Chu2025fire} against well-known datasets. 

Recently,  deepfake  detectors have encountered generalization challenges against adversarial attacks due to the lack of adversarial evaluation. Abdullah et al.~\cite{Sifat2024evolvingthreat} proposed a simple black-box attack to adversarial fake images by leveraging vision foundation models such as EfficientNet~\cite{pmlr-v97-tan19a}, ViT~\cite{dosovitskiy2021imageworth16x16words}, CLIP-ResNet~\cite{Radford2021Naturallanguage}. Their approach does not add any adversarial noise (in contrast to~\cite{eddoubi2025raiddatasettestingadversarial,Hou2023adversarial}), which might degrade the quality of an image and a viewer might identify the image as fake just by looking at it. They only adversarially manipulate the content to match it to the given text prompt; e.g., a prompt that says ``a smiling face'' should craft an adversarial fake image with a smiling face. Guided by surrogate deepfake classifiers implemented using foundation models, they adversarially updated the weights of a fake image generator, which is then used to generate adversarial fake images.
All eight detectors~\cite{Ricker2024detection, Sha2023DE-FAKE, Ojha2023ufd, Wang2024cnngenerated, chai2020makesfakeimagesdetectable, he2021Resynthesis, liu2020globaltextureenhancementfake, MesoNet2018afchar} (from 2018--2024) they tested, suffered significant performance degradation against such adversarial attacks; e.g., the worst recall degradation of 88.35$\%$  observed for DCT~\cite{Ricker2024detection} using the surrogate model CLIP-ResNet, and  a degradation of 70.85$\%$ for CNN-F~\cite{Wang2024cnngenerated}.

We evaluated adversarial attacks introduced by Abdullah et al.~\cite{Sifat2024evolvingthreat} against seven recent SOTA detectors~\cite{Karageorgiou2025anyresolution,Chu2025fire,yang2025d3,Cheng2025cospy,Brokman2025manifold,Lorch2024landscape,Tan2024upsampling} (all from 2024-2025), and found the similar performance degradation. For example, the accuracy of FIRE~\cite{Chu2025fire} on the ImageNet dataset~\cite{deng2009imagenet} dropped from 100$\%$ to 50$\%$, and the accuracy of Upsampling~\cite{Tan2024upsampling} on the DiffusionForensics~\cite{Wang2023dire} dataset dropped from 95.2$\%$ to 54.8$\%$. To avoid such severe degradations, Abdullah et al.~\cite{Sifat2024evolvingthreat} suggested the use of adversarial training as an immediate strategy to improve 
resilience. However, such adversarial 
training datasets are often scarce---leaving the SOTA 
detectors brittle against these attacks. Even more critically, the attacker can adapt and use an adversarially trained surrogate to craft a new distribution of adversarial samples, and such an adaptive attack can still significantly degrade the performance of the adversarially trained defense, making adversarial training less effective in the long-run.  
 
To counter this challenge, we propose a new approach, by  leveraging higher-order statistical modeling combined with content-agnostic feature representations and cross-scene generalization to improve robustness of SOTA detectors without requiring adversarial training. 
To date, as per our knowledge no deepfake detector has explicitly applied fourth-order statistics for GAN/diffusion image classification. We expect that fourth-order pooling can improve the performance of detectors 
as it is more sensitive to nonlinearity (compared to lower-order statistics) and strong asymmetry, 
which can potentially facilitate better generalization to new synthetic image generators. Abdullah et al.~\cite{Sifat2024evolvingthreat} showed content-agnostic features are promising to increase robustness against user-customized models. Zheng et al.~\cite{Zheng2024semanticartifacts} showed the need to break \emph{semantic artifacts} for cross-scene generalization. We combine these suggestions in our proposed approach. 

Instead of building a new detector, our goal is to show general effectiveness of the proposed methods against current adversarial attacks by integrating our approach into current SOTA detectors. Such integration is apparently feasible as evident from the six different detectors that we experimented with. We had to adjust the original network architectures of the selected detectors, in particular in $D^{3}$~\cite{yang2025d3} (due to its complex design) to incorporate our changes. Also, the integration of content-agnostic features and fourth-order statistics required high GPU resources and long training time. We conducted experiments with different order of statistics to find out which statistical order provides the desired robustness against adversarial attacks. In the end, by integrating our approach into six different types of detectors, we showed that our approach is adaptive, and could be extended to other types of deepfake detectors to improve their robustness against the adversarial attacks without the need for adversarial training.

\subhead{Main contributions}
\begin{enumerate}
\item We evaluated adversarial attacks as introduced by Abdullah et al.~\cite{Sifat2024evolvingthreat} against seven most recent SOTA detectors,  and found similar performance degradation with the maximum accuracy dropped from $98.9\%$ (for the BOSSBase steganography dataset) to $48.7\%$ for Directionality~\cite{Lorch2024landscape} (Usenix Sec'24), 
and the minimum accuracy dropped from $86.7\%$ (for out-of-domain images)
to $81.9\%$ for $D^3$~\cite{yang2025d3} (CVPR'25).

\item To enhance robustness against adversarial attacks without adversarial training, we introduce a new approach, by combining three key components: 
(i) fourth-order statistics (i.e., kurtosis) in the frequency domain to construct a frequency-based representation for classifier design and evaluation in deepfake image detection;
(ii) exploitation of directional statistics in deepfake images by applying random patch-level shuffling and rotation to improve cross-scene generalization; and
(iii) content-agnostic feature representations.
\item Our proposed method improves accuracy of the best recent SOTA detector (from our tests) $D^3$ by Yang et al.~\cite{yang2025d3}, from $81.9\%$ to $97.15\%$,  against adversarial attacks. 
When applied to the five worst performing detectors against Abdullah et al.~\cite{Sifat2024evolvingthreat}'s attacks, our approach significantly mitigated the recall degradation against such attacks, with the maximum reduced from $88.35\%$ to $9.8\%$  and the minimum reduced from $34.05\%$ to $23.3\%$. 
\end{enumerate}

\section{Related Work}
In this section, we discuss existing SOTA deepfake image detection approaches, and the recent attacks against them. 

\subsection{Existing Detection Approaches}

\subhead{Artifact-based detection methods}
These methods rely on detecting intrinsic artifacts or frequency/spatial inconsistencies introduced by diffusion or generative models. Wang et al.~\cite{Wang2023dire} proposed the DIffusion REconstruction Error (DIRE), a spatial-domain representation that measures the reconstruction error between an input image and its diffusion-based reconstruction. Inspired by DIRE, Tran et al.~\cite{Tran2025diffcor} introduced DiffCoR, a detection method that converts both real and AI-generated images into the frequency domain using the Discrete Fourier Transform (DFT). Similarly, Chu et al.~\cite{Chu2025fire} proposed Frequency-guIded Reconstruction Error (FIRE) based on the observation that diffusion models struggle to reconstruct mid-band frequency components of real images. Karageorgiou et al.~\cite{Karageorgiou2025anyresolution} investigated the spectral properties of real images and found that their spectral distributions exhibit invariant yet highly discriminative patterns. Complementarily, Ricker et al.~\cite{Ricker2024detection} analyzed the frequency characteristics of diffusion model outputs, finding that such models produce fewer high-frequency artifacts. Tan et al.~\cite{Tan2024upsampling} proposed an artifact representation termed as neighboring pixel relationships (NPR), which models second-order statistics in the spatial domain to capture local \emph{upsampling} artifacts among image pixels. Among these detectors, FIRE~\cite{Chu2025fire} demonstrated $100\%$ accuracy against well-known datasets.

\subhead{Learning-based detection methods}
Beyond artifact-level cues, several approaches leverage deep architectures to automatically learn discriminative representations of synthetic content. Yang et al.~\cite{yang2025d3} introduced the Deepfake Detector ($D^3$), a framework that incorporates an auxiliary branch trained on distorted images as discrepancy signals, thereby enhancing cross-generator generalization. Cheng et al.~\cite{Cheng2025cospy} proposed CO-SPY (COmbining Semantic and Pixel Features to Detect sYnthetic Images by AI), which integrates both semantic-level and pixel-level features for more robust synthetic image detection. Huang et al.~\cite{Huang2025sida} presented a vision-language-based framework named SIDA (Social media Image Detection, localization, and explanation Assistant) to detect tampered social media images. 
Among these detectors,  $D^3$~\cite{yang2025d3} claimed the maximum accuracy of $97.9\%$ (for in-domain images).

\subhead{Pretrained-based detection methods}
Several detectors have utilized the inherent structure of pretrained diffusion models to distinguish real from generated images. They aim to eliminate data maintenance and training altogether.
Brokman et al.~\cite{Brokman2025manifold} developed a  framework that integrates manifold analysis with diffusion model score functions. Zheng et al.~\cite{Zheng2024semanticartifacts} explored a patch-based approach designed to disrupt \emph{semantic artifacts}  in cross-scene generalization of AI-generated image detection. Cao et al.~\cite{Cao2025temporal} investigated the internal dynamics of diffusion models and identified that visual artifacts often arise during the \emph{mutation phase} of image generation, where certain spatial regions exhibit anomalous score dynamics over time; they utilized the intrinsic temporal characteristics of diffusion models themselves for more principled detection and correction of synthetic content. Among them, Brokman et al.~\cite{Brokman2025manifold} claimed the maximum accuracy of $83.9\%$. 

\subhead{Spatio-frequency statistical detection methods}
Several studies explored the inherent statistical discrepancies between real and synthetic images across both spatial and frequency domains. Corvi et al.~\cite{Corvi2023intriguing} employed second-order statistics—including the autocorrelation function and the power spectrum—to capture the characteristic deviations introduced by generative models. In a related direction, Wolter et al.~\cite{Wolter2022wavelet} proposed a multi-scale spatio-frequency representation based on wavelet-packet coefficients, which jointly captures second-order spatial and frequency statistics. These studies collectively demonstrate that deviations in natural image statistics, when analyzed across the spatial–frequency spectrum, offer powerful cues for distinguishing diffusion/GAN-generated images from real ones. Wolter et al.~\cite{Wolter2022wavelet} claimed to have an accuracy of $99.45\%$ (no results available for Corvi et al.~\cite{Corvi2023intriguing}).

\subhead{Geometry and structural inconsistency-based detection methods}
Sarkar et al.~\cite{Sarkar2024shadows} proposed a scalable, data-driven framework leveraging three complementary projective geometry cues (perspective field, linear structures and vanishing lines, and object–shadow relationships) to detect geometric discrepancies in generated imagery. Specifically, they trained geometry-aware classifiers demonstrated robustness and generalization across a wide range of generative models, including the most recent diffusion-based architectures that includes Stable Diffusion XL~\cite{Podell2024sdxl}, Kandinsky-v3~\cite{vladimir-etal-2024-kandinsky}. Complementing this perspective, Lorch et al.~\cite{Lorch2024landscape} explored the phenomenon of \emph{directionality} in images, wherein horizontal and vertical pixel sequences display distinct statistical characteristics. These findings underscore the importance of incorporating geometric and structural reasoning for building robust, rotation-invariant detectors. Sarkar et al.~\cite{Sarkar2024shadows} claimed AUC (Area Under the Curve) 0.97 for their detector. 

\subhead{Novelty of our techniques}
We identified three approaches from the existing literature and statistical methods, which can significantly improve robustness of fake image detectors against known adversarial attacks. Our approaches are designed to be integrated with existing detectors to improve their performance against existing adversarial attacks. Our methods can also be adopted selectively---one or more at a time---by detectors as suitable for their architecture and available training resources.
For example, in $D^3$~\cite{yang2025d3}, we integrated all of our techniques, and for the other five detector we used 1--2 techniques due to time and resource constraints (but still achieved significant performance improvement). 
To the best of our knowledge, no detector papers integrated the first four statistical moments, namely the mean, variance, skewness,  and kurtosis 
in the frequency domain, and thus, did not exploit this multi-order statistical pooling to expose generative traces in fake adversarial images.

\subsection{Existing Adversarial Attacks on Detectors}
In addition to detection, recent research has investigated adversarial robustness and attack strategies targeting synthetic image detectors. Abdullah et al.~\cite{Sifat2024evolvingthreat} demonstrated that lightweight customization of large generative models allows attackers to generate numerous customized fake images, significantly increasing the threat landscape. They proposed improved generalization strategies using \emph{content-agnostic} ensemble defenses that fuse foundation model features with frequency-based cues. Moreover, they showed that attackers can craft adversarial examples with \emph{semantic properties} of an image for adversarial manipulation in a fully black-box setting without explicit noise injection. Eddoubi et al.~\cite{eddoubi2025raiddatasettestingadversarial} introduced the RAID dataset, containing transferable adversarial synthetic images created using highly transferable attack strategies. They demonstrated that ensemble adversarial attacks can achieve transferability comparable to white-box settings, thereby offering a standardized benchmark for evaluating detector robustness. Finally, Hou et al.~\cite{Hou2023adversarial} proposed StatAttack, a novel natural degradation-based attack that minimizes distribution-aware loss to bridge the feature distribution gap between real and fake images. 

\begin{table}[!htb]
\setlength{\tabcolsep}{3pt}
\centering
\begin{tabularx}{0.47\textwidth}{l >{\centering\arraybackslash}X >{\centering\arraybackslash}X}
\toprule[1pt]
Defense & Acc & AP \\
\midrule[.8pt]
Brokman et al. (ManifoldBias)~\cite{Brokman2025manifold} & 0.500 & 0.706 \\
Cheng et al. (CO-SPY)~\cite{Cheng2025cospy} & 0.618 & 0.679 \\
Chu et al. (FIRE)~\cite{Chu2025fire} & 0.500 & 0.548 \\
Karageorgiou et al. (SPAI)~\cite{Karageorgiou2025anyresolution} & 0.660 & 0.678 \\
Lorch et al. (Directionality)~\cite{Lorch2024landscape} & 0.487 & 0.485 \\
Tan et al. (Upsampling)~\cite{Tan2024upsampling} & 0.548 & 0.569 \\
Yang et al. ($D^3$)~\cite{yang2025d3} & \textbf{0.819} & \textbf{0.909} \\
\bottomrule[1pt]
\end{tabularx}
\caption{Performance of recent SOTA detectors tested against the Surrogate StyleCLIP dataset~\cite{Sifat2024evolvingthreat} in terms of Accuracy (Acc) and Average Precision (AP)}
\label{Tab:Detector_SOTA}
\end{table}

\begin{table}[!htb]
\footnotesize
\setlength{\tabcolsep}{3.5pt} 
\smallskip
\centering
\begin{tabularx}{0.47\textwidth}{llcc}
\toprule[1pt]
Attack & Dataset & Acc & AP \\
\midrule[.8pt]
Sec3 & SDdataset & 0.820 & 0.884  \\
Sec3 & StyleCLIP & 0.808 & 0.896  \\
Sec5.1 FM genimages & AbsoluteReality & 0.873 & 0.923  \\
Sec5.1 FM genimages & AnalogDiffusion & 0.800 & 0.860  \\
Sec5.1 FM genimages & DreamlikePhotoreal & 0.871 & 0.933  \\
Sec5.1 FM genimages & DreamShaper & 0.893 & 0.948  \\
Sec5.1 FM genimages & EpicDiffusion & 0.842 & 0.903  \\
Sec5.1 FM genimages & epiCRealism & 0.831 & 0.891  \\
Sec5.1 FM genimages & FotoAssistedDiffusion & 0.839 & 0.901  \\
Sec5.1 FM genimages & Haveall & 0.867 & 0.923  \\
Sec5.1 LoRA genimages & AddAestheticDetails & 0.765 & 0.839  \\
Sec5.1 LoRA genimages & AddDetails & 0.765 & 0.833  \\
Sec5.1 LoRA genimages & EnhanceImage & 0.765 & 0.842  \\
Sec5.1 LoRA genimages & ImproveContrast & 0.826 & 0.897 \\
Sec5.1 LoRA genimages & IncreaseSharpness & 0.686 & 0.756  \\
Sec5.1 LoRA genimages & ReduceImageNoise & 0.773 & 0.845  \\
Sec5.1 LoRA genimages & TweakBrightness & 0.771 & 0.798  \\
Sec5.1 LoRA genimages & TweakSkinTexture & 0.758 & 0.826  \\
Sec5.2 AdvImageswSurrogate & CLIPResNet & 0.788 & 0.854  \\
Sec5.2 AdvImageswSurrogate & EfficientNet & 0.859 & 0.923  \\
Sec5.2 AdvImageswSurrogate & ViT & 0.813 & 0.902  \\
Sec5.2 AdvtrainedCLIPResNet & advV2 & 0.752 & 0.832 \\
Sec5.2 Dataset for Adv. training & ForDefenses & 0.778 & 0.841 \\
Sec5.2 Dataset for Adv. training & ForSurrogate & 0.796 & 0.855 \\
Sec5.2 Surrogate-StyleCLIP & StyleCLIP & 0.819 & 0.909 \\
\bottomrule[1pt]
\end{tabularx}
\caption{Performance of $\bm{D}^3$ against all datasets of Abdullah et al.~\cite{Sifat2024evolvingthreat} in terms of  Accuracy and Average Precision}
\label{Tab:D3Detector}
\end{table}

\section{Evaluation of Recent SOTA Detectors against New Attacks}
\label{sec:sota-eval}
Several new detectors in top academic venues came out after the publication of the attacks by Abdullah et al.~\cite{Sifat2024evolvingthreat}. However, these new detector papers do not provide any evaluation against these known attacks. 
To understand the effectiveness of such attacks against the newly designed detectors, we chose seven recent (published between 2024-2025) SOTA  detectors~\cite{Karageorgiou2025anyresolution,Chu2025fire,yang2025d3,Cheng2025cospy,Brokman2025manifold,Lorch2024landscape,Tan2024upsampling}, according to the following criteria. 
(1) Detection performance: we prioritized defenses that reported strong detection results.
(2) Code availability: only models with publicly available checkpoints and training code were considered. 
(3) Content diversity: beyond standard face images, we included defenses evaluated on varied content types, such as artwork, illustrations, and images of different classes of objects. (4) Methodological variety: the selected defenses employ diverse approaches, allowing us to assess the robustness of different detection strategies comprehensively.
We did not perform any fine-tuning on these seven models; rather, we utilized their officially published weights to evaluate them against the adversarial attack of Abdullah et al.~\cite{Sifat2024evolvingthreat}. The results were compared using accuracy (ACC) and average precision (AP), consistent with the metrics employed in the original publications.
Note that
we excluded the attacks from Eddoubi et al.~\cite{eddoubi2025raiddatasettestingadversarial} and Hou et al.~\cite{Hou2023adversarial} as their adversarial images are generated by adding noise, which may degrade the image quality significantly (see Sec.~\ref{sec:datasets}, under ``Evaluating the quality of adversarial data'').

Evaluating the SOTA detectors against the selected attacks required non-trivial efforts. A common issue with most of detectors' code is mismatched machine learning library versions causing the code to crash. 
We also had to make small changes to make the detectors run in our environments. 
In all cases, we first reproduced the published results of the tested detector before performing tests with the adversarial datasets.

Our results are summarized in Table~\ref{Tab:Detector_SOTA}. It is evident that D$^3$~\cite{yang2025d3} is the best detector against the selected adversarial attacks, while all other detectors performed very badly, with some detectors' results became no better than random guessing (accuracy $\leq$ 0.5). To understand the performance of D$^3$ more thoroughly, we evaluated it against all types of test datasets provided by Abdullah et al.~\cite{Sifat2024evolvingthreat}, including user-customized generative models, and  datasets generated by controlling content and image quality; see Table~\ref{Tab:D3Detector} for the results. Evidently, for all datasets, D$^3$  shows performance degradation, including the first two test datasets (labeled Sec3)  which address the issue of controlling the content and image quality. It shows that performance of SOTA detectors might degrade against high quality fake test data, if the training fake dataset quality is not maintained, which is consistent with the findings of Abdullah et al. They also emphasized for generalization of the defenses against user-customized models by demonstrating performance degradation of their selected defenses tested against such models. In Table~\ref{Tab:D3Detector}, under all the rows labeled Sec5.1, we also demonstrated performance degradation against two approaches -- Full Model fine-tuning (FM~\cite{li2020rethinking}) and Low-Rank Adaptation (LoRA~\cite{hu2022lora}) used by Abdullah et al. The rest of the results labeled Sec5.2 in Table~\ref{Tab:D3Detector} shows performance degradation against different adversarial attacks.

Overall, we reaffirmed the significant performance degradation similar to Abdullah et al.~\cite{Sifat2024evolvingthreat} for the most recent SOTA detectors, and even the best performing detector (D$^3$), is affected by all of the test datasets. Such results highlight the need for robust new mechanisms to improve the detectors.

\begin{figure*}[!htb]
\centering
\includegraphics[height=.8\textwidth, angle=90]{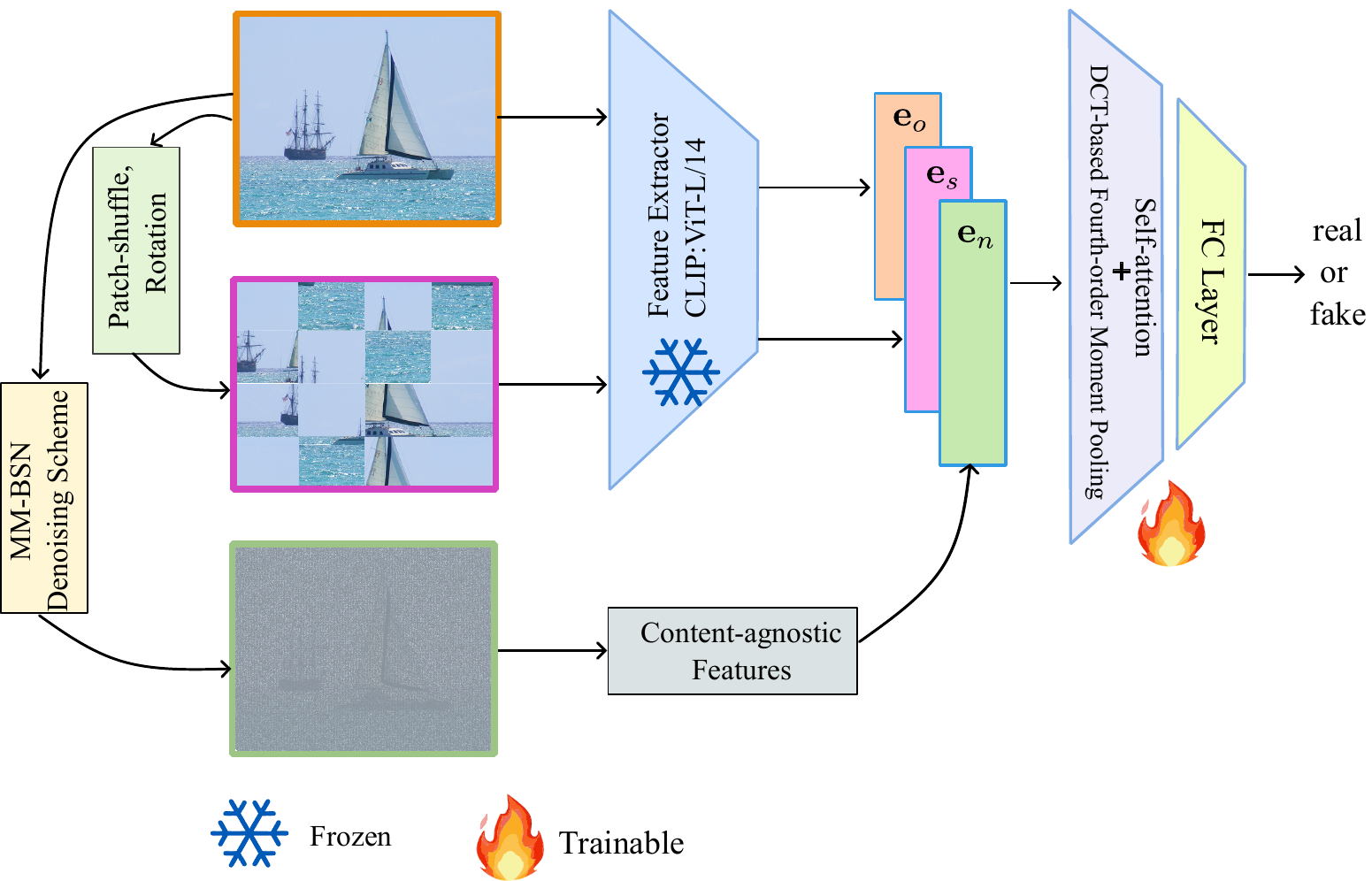}
\caption{Overall framework of our enhanced deepfake detector as integrated with $\bm{D}^3$. The original image and its patch-shuffled variant with our new addition of random rotation  are fed into the pretrained vision model for feature extraction. We add a new MM-BSN denoising scheme as the third branch to extract content-agnostic features from the same input image. We then utilize the self-attention module (as in $\mathbf{D}^\mathbf{3}$) to encourage the learning of shared artifacts between the features, which are then concatenated with our newly added frequency-aware fourth-order statistics captured with DCT-based fourth-order moment pooling. Finally, we use a single, fully-connected (FC) layer linear classifier to get the final prediction (as in $\bm{D}^3$).}
\label{Fig:network_architecture}
\end{figure*}

\section{Methodology}
In this section, we present a detector that attains robustness to content‑manipulating adversaries without adversarial training.

\subsection{Problem Formulation and Threat Model}
Let $\mathcal{G} = \{G_i\}_{i=1}^N$ be a generator set comprised of $N$ different types of generators, and denote by $\mathcal{R}_i = \{r_1^{(i)}, r_2^{(i)}, \dots, r_{N_i}^{(i)}\}$ and $\mathcal{F}_i = \{f_1^{(i)}, f_2^{(i)}, \dots, f_{N_i}^{(i)}\}$ the real and fake samples of $G_i$, respectively, where $N_i$ is the number of samples in each class. Our objective is to correctly distinguish between real and fake samples when presented with any image from the overall dataset $\mathcal{D}= \{\{\mathcal{R}_i\}_{i=1}^N \cup \{\mathcal{F}_i\}_{i=1}^N\}$.

For the threat model, we assume attackers are interested in using fully synthetic, high-quality deepfake images for malicious purposes. They have the resources to use open-source SOTA image generators by themselves, or can access commercial generators. They also exploit the best known adversarial techniques in their generated images to defeat the current deepfake detectors. We exclude partially modified images, which can also be used for malicious purposes. 

\subsection{Guiding Principles}
We adhere to the following guiding principles in our design: 

\subhead{Disentangle semantic cues from generative traces} To develop a universal detector for AI-generated images, it is essential to consider not only generator artifacts but also the semantic artifacts inherent in visual content. Generator artifacts arise from the synthetic image generation process itself, whereas semantic artifacts emerge from the semantic composition of images, or from biases introduced during dataset curation and preprocessing. Notably, semantic artifacts exhibit properties analogous to generator artifacts and may inadvertently be exploited by detection models. 

\subhead{Prioritize content‑agnostic evidence} To enhance generalization across diverse scenes, it is crucial to mitigate the overfitting of semantic artifacts during training~\cite{Zheng2024semanticartifacts}. Moreover, effective defenses should prioritize intrinsic imperfections in synthesized images rather than being influenced by shifts in content distribution. Consequently, incorporating content-agnostic features into detection frameworks is vital for improving robustness and cross-domain generalization~\cite{Sifat2024evolvingthreat}.

\subsection{Proposed Framework}

\subsubsection{Architecture Overview}
The overall architecture of the proposed framework is illustrated in Figure~\ref{Fig:network_architecture} (integrated with $\bm{D}^3$). Conventional learning‑based detection approaches often inadequately account for intrinsic characteristics of authentic images, which can lead to overfitting and limited generalization performance, particularly under distribution shifts~\cite{yang2025d3}. To address this limitation, we adopt a pretrained‑based detection paradigm, wherein feature representations are extracted from a model not explicitly optimized for generated image detection. Such representations benefit from more unbiased decision boundaries and improved robustness across diverse image domains. 

To enhance cross‑scene generalization and mitigate the influence of semantic \emph{overfitting}, we disrupt global semantic structures while preserving localized artifact signatures. We partition the input image into non‑overlapping patches and apply random transformations (shuffling and rotation), so that the model does not rely on high-level scene context (i.e., the semantic arrangement of objects). This forces the detector to extract universal, low‑level features (e.g., DCT kurtosis, pixel gradient magnitudes, color covariance) that are invariant to semantic content.  
Here, high-level scene context refers to the global structural coherence and semantic relationships between objects, such as facial geometry, anatomical symmetry, and environmental spatial logic.
Low-level features are defined as the stationary statistical properties of the image signal that persist regardless of spatial orientation---including frequency-domain distributions (e.g., DCT kurtosis), first-order spatial statistics (e.g., pixel gradient magnitudes), and inter-channel spectral dependencies (e.g., color covariance).

Prior studies have also demonstrated that the noise residual (or noise space) of an image, obtained by suppressing semantic content, can encode discriminative signatures valuable for deepfake detection~\cite{Jiameng2020noisescope}. Building upon this observation, we employ a state‑of‑the‑art denoising method, MM‑BSN~\cite{Zhang2023mmbsn}, to extract noise residuals for all training images. Features extracted from these residuals serve as content‑agnostic representations, further strengthening generalization by reducing sensitivity to image semantics. Within our architecture, the original input image, along with its patch‑shuffled, and rotated variant, is fed into a pretrained CLIP ViT‑L/14 model~\cite{Radford2021Naturallanguage} to obtain visual embeddings. Features are extracted from the penultimate layer of the CLIP network, as this layer preserves richer and more detailed visual information compared to the final embedding layer. In parallel, content‑agnostic features are extracted from the corresponding noise residuals of the same input image. Subsequently, a self‑attention mechanism, together with frequency-aware fourth-order statistics, is employed to enable effective interaction and discrepancy learning among the three feature modalities. This design allows the model to emphasize inconsistencies indicative of synthesized content while suppressing semantic bias. Finally, a linear classifier aggregates the fused representations to predict binary real or fake labels. 

\subsubsection{Framework Components} 
We now provide the details of the components of our architecture (see Figure~\ref{Fig:network_architecture}).

\subhead{Triple-path discrepancy learning} To encourage the model to capture deeper and more robust artifacts, we introduce two additional branches in conjunction with the primary input pathway. Specifically, the first branch processes the original input image, the second branch operates on a patch-shuffled and rotated variant of the same image, and the third branch incorporates content-agnostic features derived from the corresponding noise residual. This multi-branch design is intended to reduce reliance on semantic cues while promoting the learning of invariant and generalizable discriminative patterns. 

Formally, let an original image $\bm{X}_o$ be sampled from the source dataset $\mathcal{D}_s = \Bigl\{ \{R_i\}_{i=1}^{k} \cup \{F_i\}_{i=1}^{k} \Bigr\}$, where $R_i$ and $F_i$ denote real and synthesized images, respectively, and $\bm{X}_o \in \mathcal{G}_s \subset \mathcal{D}_s$. Two visual embeddings are extracted using a pretrained CLIP backbone. The first embedding is obtained from the augmented original image as
\begin{equation}
	\bm{e}_o = \mathrm{CLIP}^*\bigl(\mathrm{AUG}(\bm{X}_o)\bigr).
\end{equation}

Here, the operator $\mathrm{AUG}(\cdot)$ represents standard data augmentation strategies commonly adopted for generated image detection, such as random cropping, resizing, and color jittering~\cite{Wang2024cnngenerated}.
The second embedding is generated from a patch-shuffled version of the image, which suppresses superficial and semantic artifacts:
\begin{equation}
	\bm{e}_s = \mathrm{CLIP}^*\bigl(\mathrm{PS}(\mathrm{AUG}(\bm{X}_o))\bigr).
\end{equation}

In the patch-shuffling operation $\mathrm{PS}(\cdot)$, the image is divided into fixed-size patches that are randomly permuted and independently subjected to random rotations (e.g., $90^\circ$, $180^\circ$, or $270^\circ$). This operation effectively disrupts global spatial structure while preserving local statistics. The embeddings $\bm{e}_o$ and $\bm{e}_s$ are $D$-dimensional visual feature vectors extracted from the penultimate layer of the pretrained CLIP model, denoted by $\mathrm{CLIP}^*$, which retains richer visual representations than the final projection layer. 

\subhead{Content‑agnostic residual extraction}
We employ the multi‑mask blind‑spot network (MM‑BSN)~\cite{Zhang2023mmbsn} to estimate a noise residual map from a noisy observation $\bm{X}_o$. MM-BSN was selected over alternatives (e.g., DnCNN~\cite{zhang2017beyond} and  Blind2Unblind~\cite{Wang2022Blind2Unblind}) as it operates in a self-supervised manner, eliminating the need for clean reference images during training, and it has demonstrated state-of-the-art performance on real-world noise estimation benchmarks. To effectively disrupt spatial dependencies in the noise while preserving fine-grained texture information, MM-BSN adopts a multi-mask strategy that fuses features extracted from multiple masked convolutional kernels. Following the multi-branch feature aggregation paradigm introduced by Sun et al.~\cite{Sun2019HRNet}, features from parallel masked pathways are jointly integrated to predict a spatial noise residual map~$\hat{\bm{V}}\in \mathbb{R}^{H \times W \times 3}$ as follows:
\begin{equation}
\hat{\bm{V}} = \mathcal{F}_{\mathrm{fuse}}\bigl(\mathrm{Concat}(f_{m_1}(\bm{X}_o), f_{m_2}(\bm{X}_o), \dots, f_{m_k}(\bm{X}_o)))\bigr) ,
\end{equation}
where $f_{m_i}(\cdot)$ denotes the $i$-th masked processing branch. The hierarchical fusion network $\mathcal{F}_{\text{fuse}}$ comprises concatenation-based dilated convolutional layers (CDCLs) that aggregate multi-scale spatial context. By incorporating a limited number of dilated convolutional layers (DCLs) within each CDCL block, the effective receptive field is expanded to capture surrounding contextual information while avoiding the parameter growth associated with naive stacking of standard convolutions. This design facilitates efficient interaction among different mask pathways, thereby reinforcing texture recovery while maintaining a compact model footprint.

Unlike feature representations extracted directly from the original image, the noise artifacts require a compact yet expressive encoding. Inspired by the projection head design used by Chen et al.~\cite{Chen2020SimCLR}, we introduce a noise projection module $\mathcal{P}_{\text{noise}}$ to map the spatial residual features into a low-dimensional latent space. Specifically, the noise embedding $\bm{e}_n \in \mathbb{R}^{D}$ is defined as
\begin{equation}
	\bm{e}_n = \mathcal{P}_{\text{noise}}(\hat{\bm{V}}),
\end{equation}
where $\mathcal{P}_{\mathrm{noise}}(\cdot) = \mathrm{MLP}(\text{GAP}(\cdot))$. Here, $\mathrm{GAP}$ denotes Global Average Pooling, which collapses the spatial dimensions $H \times W$ into a channel-wise mean vector, and $\mathrm{MLP}$ represents a multi-layer perceptron that projects the resulting vector into a $D$-dimensional embedding space. Consequently, $\bm{e}_n$ encodes global noise characteristics as a compact feature vector, matching the dimensionality of the CLIP-based embeddings $\bm{e}_o$ and $\bm{e}_s$.

This tri-path representation is designed to enhance robustness and generalization to unseen generators and post-processing manipulations. Building upon the complementary embeddings $\bm{e}_o$, $\bm{e}_s$, and $\bm{e}_n$, the subsequent objective is to integrate their heterogeneous feature representations into a unified invariant space. To this end, we introduce a dual-branch fusion module that jointly exploits self-attention mechanisms and fourth-order moment pooling, enabling the model to capture both relational dependencies and higher-order distributional consistencies across the three feature pathways.

\subhead{Self-attention and 4th-order invariant feature extraction} Following the triple-path design described above, each input image $\bm{X}_o$ gives rise to three complementary feature representations: a semantic-preserving embedding from the original image branch $\bm{e}_o$, an artifact-sensitive embedding from the patch-shuffled branch $\bm{e}_s$, and a content-agnostic noise-based embedding $\bm{e}_n$ obtained via MM-BSN-based denoising and residual extraction. All three representations are extracted from the penultimate layer of the pretrained CLIP backbone and are projected into a common feature space of dimensionality $D$. Consequently, $\bm{e}_o$, $\bm{e}_s$, and $\bm{e}_n$ are vectors in $\mathbb{R}^{D}$ (with $D=1024$), enabling direct comparison and joint modeling. These embeddings are then stacked and forwarded to the attention and pooling modules for feature fusion. Specifically, the three branch-specific representations are concatenated along the branch dimension to form a unified feature tensor:
\begin{equation}
	\bm{E} = \mathrm{Concat}(\bm{e}_o, \bm{e}_s, \bm{e}_n) \in \mathbb{R}^{D \times 3}.
\end{equation}
To capture higher-order invariant statistics across branches, a linear dimensionality reduction operator $\mathcal{R}: \mathbb{R}^{D} \rightarrow \mathbb{R}^{D/4}$ is first applied independently to each branch embedding. Note that the reduction from $D=1024$ to $D/4=256$ before DCT-based pooling serves two purposes: it reduces the memory footprint of the fourth-order moment computation (which scales as $O(D^2)$ in the covariance terms) to a tractable size, and it acts as a mild bottleneck that encourages the representation to retain only the most statistically stable features before moment computation. The reduced representation is then processed by a \emph{DCT-based fourth-order moment pooling} module, which performs a DCT-II transform along the branch dimension and computes the first four statistical moments, namely the mean, variance, skewness, and kurtosis, of the transformed coefficients:
\begin{equation}
	\tilde{\bm{E}} = \mathcal{R}(\bm{E}), \quad
	\bm{f}_{4} = \mathcal{P}_{4}^{\mathrm{DCT}}(\tilde{\bm{E}}) \in \mathbb{R}^{D}.
\end{equation}
This operation encodes frequency-aware fourth-order statistics while remaining invariant to branch ordering, thereby enhancing robustness to viewpoint and representation perturbations.
In parallel, the original stacked feature tensor $\bm{E}$ is fed into a Transformer-based self-attention module that operates along the branch dimension:
\begin{equation}
	\bm{f}_{\mathrm{attn}} = \mathcal{A}(\bm{E}) \in \mathbb{R}^{D},
\end{equation}
where attention weights are learned to model cross-branch dependencies and enforce consistency among the original, patch-shuffled, and noise-derived views.
The resulting descriptors are $\ell_2$-normalized, yielding $\hat{\bm{f}}_{\mathrm{attn}}$ and $\hat{\bm{f}}_{4}$, and then concatenated to construct the final multi-view invariant representation:
\begin{equation}
	\bm{f}_{\mathrm{fused}} = \mathrm{Concat}(\hat{\bm{f}}_{\mathrm{attn}}, \hat{\bm{f}}_{4}).
\end{equation}
Intuitively, the original branch embedding $\bm{e}_o$ preserves semantic fidelity, the patch-shuffled embedding $\bm{e}_s$ enforces robustness by disrupting spatially correlated shortcuts, and the noise-based embedding $\bm{e}_n$ introduces content-independent statistical cues that emphasize generator-specific artifacts. By jointly modeling transformer-based self-attention and DCT-based 4th-order statistics, the proposed fusion strategy achieves invariance to spatial perturbations while remaining sensitive to subtle statistical traces characteristic of synthesized imagery.

\subhead{Classification head} 
The fused representation is passed through a linear classification layer to produce the final prediction:
\begin{equation}
	\hat{\bm{y}} = \sigma(\bm{W}\bm{f}_{\mathrm{fused}} + \bm{b}),
\end{equation}
where $\sigma(\cdot)$ denotes the sigmoid activation function, $\bm{W}$ is the learnable weight matrix of the fully-connected (FC) layer, and $\bm{b}$ is the bias term. The model parameters of the classifier are optimized using the binary cross-entropy loss:
\begin{equation}
\mathcal{L}=-\frac{1}{M}\sum_{i=1}^{M}\left[y_i \log \hat{y}_i +(1 - y_i)\log \big(1 - \hat{y}_i \big)\right],
\end{equation}
where $M$ is the batch size, $y_i \in \{0,1\}$ is the ground-truth label of the $i$-th sample, and $\hat{y}_i$ is the prediction. During training, the CLIP backbone remains frozen, while the self-attention module and classification head are optimized, ensuring that learning focuses on invariant relational and higher-order statistical representations. This design plugs into discrepancy‑based frameworks (e.g., $\bm{D}^3$) with minimal changes and avoids adversarial training, yet targets the very weaknesses exposed by black‑box surrogate attacks.

\medskip\noindent\emph{Why the combination improves robustness?} Patch shuffle and rotations suppress scene semantics and orientation cues responsible for cross‑scene failures, compelling reliance on universal, low‑level regularities; MM‑BSN contributes content‑agnostic evidence; and fourth‑order frequency pooling emphasizes heavy‑tailed spectral irregularities harder to steer via surrogate‑guided semantic editing. Together, these modules directly counteract the vulnerabilities highlighted in recent robustness studies while remaining training‑ and architecture‑agnostic~\cite{Zheng2024semanticartifacts,Zhang2023mmbsn}.

\section{Experiments and Results}

\subsection{Datasets, Quality, and Evaluation Metrics}\label{sec:datasets}

\subhead{Datasets for training} 
To assess the performance of our proposed framework, initially we considered experiments on both the UniversalFakeDetect (UFD)~\cite{Ojha2023ufd} and GenImage~\cite{Zhu2024genimage} datasets. The UFD dataset comprises approximately 720K training images, evenly split between 360K authentic samples sourced from LSUN~\cite{yu2015lsun}, LAION-400M~\cite{schuhmann2021laion400m}, and ImageNet~\cite{deng2009imagenet}; and 360K synthetic images produced by a diverse set of generative models (e.g., ProGAN~\cite{karras2018progressive}, StyleGAN~\cite{karras2019stylegan}, 
GLIDE~\cite{nichol2022glide}, and DALL·E~\cite{ramesh2021zeroshot}).
However, later we considered GenImage~\cite{Zhu2024genimage} dataset only for our training and evaluation as it is more recent, and robust in terms of quality, number of images, types of generators, and content variations. The GenImage dataset consists of approximately 2.68 million samples, divided into 1.33 million real and 1.35 million synthetic images. For each generator type, 50,000 images are reserved for testing. The synthetic subset is produced by eight different generative models, including one GAN (BigGAN~\cite{brock2019biggan}) and seven diffusion-based models: Stable Diffusion V1.4~\cite{rombach2022ldm}, Stable Diffusion V1.5, GLIDE~\cite{nichol2022glide}, VQDM~\cite{gu2022vqdm}, Wukong~\cite{gu2022wukong}, ADM~\cite{dhariwal2021adm}, and Midjourney~\cite{midjourney2022}. 

We extended the GenImage dataset by adding StyleCLIP datasets (labeled as Sec3, non-adversarial data) from Abdullah et al.~\cite{Sifat2024evolvingthreat} as a separate class, which has a balanced set of 16,000, 2,000 and 2,000 images across both fake and real classes for training, validation, and testing, respectively. For training the modified $D^3$ model with our methodology, we used only 36,000 images from the extended GenImage dataset (due to time and resource constraints), randomly selected in equal number from each of its classes, and for validation we used all the given data.  
We maintained content variations even though selecting only a limited number of images for training (36,000 out of 2.68 million). Importantly, we did not use any adversarial data for training.

\subhead{Datasets for testing}
To test our trained models against adversarial attacks, initially we considered using two datasets RAID~\cite{eddoubi2025raiddatasettestingadversarial} and Abdullah et al.~\cite{Sifat2024evolvingthreat}. 
RAID provides approximately 72,000 adversarially-perturbed images derived from multiple text-to-image models, designed to systematically test how well image-generation detectors withstand deliberate evasion attempts. The dataset is constructed by attacking an ensemble of seven SOTA detectors, and includes examples that transfer across unseen detectors, thereby revealing significant vulnerabilities in current detection systems~\cite{eddoubi2025raiddatasettestingadversarial}. 

Abdullah et al.~\cite{Sifat2024evolvingthreat} introduced a dataset of \emph{adversarial fake images} generated by leveraging large vision foundation models to perform \emph{content-aware} manipulations rather than classic pixel-level perturbations.  Given an already deceptive synthetic image, the attacker specifies a semantic attribute via a text prompt (e.g., ``a smiling face''), and uses the foundation model to alter that attribute while preserving the image's realism.  Crucially, this procedure does not rely on additive adversarial noise; instead, it modifies high-level semantics to evade detectors that were trained on non-semantic corruptions (cf.\ RAID's noise-based adversarial perturbations).  Empirical evaluation~\cite{Sifat2024evolvingthreat} shows that these content-aware, semantically guided edits significantly degrade detector performance. 

\subhead{Evaluating the quality of adversarial data}
Abdullah et al.~\cite{Sifat2024evolvingthreat} used the Kernel Inception Distance (KID)~\cite{kid-test} metric to measure synthetic image quality. KID measures the distribution distance between the real and fake image sets. The values are unbounded, and smaller values (closer to zero) indicate better synthetic image quality, i.e., fake images match the distribution of the real images.
Abdullah et al.\ obtained a KID score of 0.008 from their SD dataset. For the RAID~\cite{eddoubi2025raiddatasettestingadversarial} dataset, we computed the KID  score of 0.026, which is far larger (over 3.25 times compared to~\cite{Sifat2024evolvingthreat}), indicating degraded image quality (which we also manually verified). So, we finally excluded the RAID dataset for the evaluation of our models and the baselines. We also excluded the adversarial noise (perturbations) based StatAttack~\cite{Hou2023adversarial} due to the same reason (also as their dataset is not public, we could not compute the corresponding KID score).

\subhead{Evaluation metrics}
We employ classification accuracy and average precision (AP) to evaluate the performance of the selected detectors, using a fixed threshold of 0.5 when calculating the accuracy score for each detector, following $D^3$~\cite{yang2025d3}.
AP measures how well a model distinguishes between real and synthetic images, independent of any particular decision threshold. Some prior studies~\cite{Wang2024cnngenerated,Wang2023dire,Ojha2023ufd,liu2022generatedimages} often report the mean Average Precision (mAP) by computing the AP for each image generator separately and then taking their average. However, this simple averaging approach overlooks the considerable variability in optimal classification thresholds across different generators~\cite{everingham2010pascal,Wang2024cnngenerated}. In realistic deployment scenarios, users cannot feasibly fine-tune detection thresholds for every unseen generator. We thus adopt a unified evaluation metric by computing a global AP that enforces consistent decision standards across all tested generators, following $D^3$~\cite{yang2025d3}. 

\subsection{Implementation and Results}
\label{sec:impl-results}
In this section we provide  implementation details to incorporate our changes to the six target detectors, and results of our proposed improvement. We first discuss $D^3$~\cite{yang2025d3}, which has been selected among the seven recent SOTA detectors for its comparatively best performance (see Sec.~\ref{sec:sota-eval}) against adversarial attacks. Then we discuss the five worst performing detectors selected from the eight detectors of Abdullah et al.~\cite{Sifat2024evolvingthreat} for improvement by implementing our techniques.

\subhead{Improvement of $\mathbf{D}^\mathbf{3}$}
Following its original design~\cite{yang2025d3}, in our experiments, we utilize the CLIP ViT-L/14 model, pretrained on the extensive WebImageText dataset, as the backbone for extracting patch-level representations. To preserve rich information, we use the penultimate layer features prior to any dimensionality reduction. With the backbone frozen, the whole network is trained on a single NVIDIA A100 80GB GPU with a batch size of 64, 
except for the experiment with fourth-order statistics, for which we used a H100 96GB GPU due to more GPU memory demand.  During training, we employ the Adam optimizer~\cite{Kingma2014AdamAM} with an initial learning rate of \(1 \times 10^{-4}\) and no weight decay. 
All input images are first resized to \(256 \times 256\) and subsequently randomly cropped to \(224 \times 224\)(native resolution of the backbone model CLIP: ViT-L/14). 
The data augmentation strategy follows prior work~\cite{Ojha2023ufd, Wang2024cnngenerated}, incorporating Gaussian blur and JPEG compression to enhance robustness. Specifically, the JPEG compression quality is uniformly sampled within the range of 30 to 100, while the Gaussian blur parameter \(\sigma\) is randomly selected from the interval \([0, 3]\), each applied with a probability of 0.5. 
During validation and testing, all images are resized to \(224 \times 224\) without applying any data augmentation to ensure the evaluation is deterministic. For training the modified $D^3$ up to third-order pooling, it took around 24 hours per epoch in an A100 80GB GPU; for fourth-order pooling it took around 14 hours per epoch in a H100 96GB GPU.

\begin{table*}[!htb]

\setlength{\tabcolsep}{6pt}
\centering
\begin{tabularx}{\textwidth}{
l
>{\raggedright\arraybackslash}X
>{\centering\arraybackslash}X
>{\centering\arraybackslash}X
>{\centering\arraybackslash}X
>{\centering\arraybackslash}X
}
\toprule[1pt]
Attack & Dataset &
\multicolumn{2}{c}{Acc} &
\multicolumn{2}{c}{AP} \\
\cmidrule(lr){3-4} \cmidrule(lr){5-6}
& & $D^3$ & Ours & $D^3$ & Ours \\
\midrule[.8pt]
Sec5.2 AdvImages w/ Surrogate Models & CLIPResNet & 0.788 & 0.9120 & 0.854 & 0.9466 \\
Sec5.2 AdvImages w/ Surrogate Models & EfficientNet & 0.859 & 0.9480 & 0.923 & 0.9673 \\
Sec5.2 AdvImages w/ Surrogate Models & ViT & 0.813 & 0.9285 & 0.902 & 0.9635 \\

Sec5.2 Advtrained CLIPResNet & advV2 & 0.752 & 0.9230 & 0.832 & 0.9549 \\
Sec5.2 Dataset for Adv. training & ForDefenses & 0.778 & 0.9245 & 0.841 & 0.9540 \\
Sec5.2 Dataset for Adv. training & ForSurrogate & 0.796 & 0.9275 & 0.855 & 0.9506 \\
Sec5.2 surrogate-StyleCLIP-dataset & StyleCLIP & 0.819 & 0.9715  & 0.909 & 0.9795  \\

\bottomrule[1pt]
\end{tabularx}
\caption{Performance of our model vs.\ the original $D^3$ defense, tested against adversarial datasets (Abdullah et al.)}
\label{Tab:D3VsOursDetector}
\end{table*}

As summarized in Table~\ref{Tab:D3VsOursDetector}, our proposed modifications in $D^3$ consistently outperform the original model across all evaluated adversarial datasets. Specifically, we achieved a substantial improvement in classification accuracy, with an average gain of $+16.72\%$ over the baseline. Surprisingly, the most significant enhancement is observed for the \textit{advV2} dataset, which is created by Abdullah et al.\ using the most effective surrogate model shown in their work (CLIP-ResNet),  where our model yields a $+22.74\%$ increase in accuracy, demonstrating superior robustness against CLIP-based adversarial perturbations. Furthermore, the consistent gains in Average Precision (AP), averaging $+9.94\%$, indicate that our model maintains a more reliable ranking of fake samples even under sophisticated surrogate-model attacks. These results suggest that by integrating fourth-order frequency statistics with self-attentive cross-branch modeling, our architecture successfully mitigate adversarial attack without adversarial training that remain elusive to recent SOTA defense mechanisms.

\begin{table*}[!htb]

\small
\centering
\begin{tabularx}{\textwidth}{
    l 
    *{3}{>{\centering\arraybackslash}X}   
    *{4}{>{\centering\arraybackslash}X}   
    *{3}{>{\centering\arraybackslash}X}   
    *{4}{>{\centering\arraybackslash}X}   
    *{3}{>{\centering\arraybackslash}X}   
}
\toprule[1pt]

Surrogate &
\multicolumn{3}{c}{DCT~\cite{Ricker2024detection}} &
\multicolumn{4}{c}{DE-FAKE~\cite{Sha2023DE-FAKE}} &
\multicolumn{3}{c}{CNN-F~\cite{Wang2024cnngenerated}} &
\multicolumn{4}{c}{Patch-Forensics \cite{chai2020makesfakeimagesdetectable}} &
\multicolumn{3}{c}{Resynthesis~\cite{he2021Resynthesis}} \\
\cmidrule(lr){2-4} \cmidrule(lr){5-8} \cmidrule(lr){9-11} \cmidrule(lr){12-15} \cmidrule(lr){16-18}

Model & SP24 & Ours & Imp. 
      & SP24 & Org.\ & Ours & Imp. 
      & SP24 & Ours & Imp. 
      & SP24 & Org.\ & Ours & Imp. 
      & SP24 & Ours & Imp. \\
\midrule[.8pt]

EfficientNet &
57.43 & 27.81 & \textbf{51.6\%} &
75.76 & 26.90 & 3.63 & \textbf{86.5\%} &
40.06 & 15.44 & \textbf{61.5\%} &
28.37 & 43.43 & 30.89 & \textbf{28.9\%} &
44.58 & 23.21 & \textbf{47.9\%} \\

ViT &
53.41 & 16.76 & \textbf{68.6\%} &
78.43 & 28.57 & 8.48 & \textbf{70.3\%} &
32.23 & 10.50 & \textbf{67.4\%} &
13.47 & 34.05 & 23.30 & \textbf{31.6\%} &
37.39 & 16.22 & \textbf{56.6\%} \\

CLIP-ResNet&
88.35 & 9.80 & \textbf{88.9\%} &
80.04 & 40.49 & 11.63 & \textbf{71.3\%} &
70.85 & 15.03 & \textbf{78.8\%} &
40.96 & 72.76 & 58.60 & \textbf{19.5\%} &
73.96 & 24.93 & \textbf{66.3\%} \\

\bottomrule[1pt]
\end{tabularx}
\caption{Evaluation of the defenses on adversarial fake images created using Abdullah et al.~\cite{Sifat2024evolvingthreat} (labeled as SP24). We use ``Org.'' for our reproduction of SP24 attacks, where our results significantly differ with SP24. We report $\Delta R$ for the fake class in $\%$, where lower is better. The ``Imp.'' columns show the percentage reduction in recall degradation achieved by our modified model relative to the baseline.}
\label{Tab:DetectorTransposed}

\end{table*}

\subhead{Improvement of weak detectors}
We selected the five worst-performing detectors from  Abdullah et al.~\cite{Sifat2024evolvingthreat} to test if those could be improved by implementing our proposal. The five detectors are: DCT~\cite{Ricker2024detection}, DE-FAKE~\cite{Sha2023DE-FAKE}, CNN-F~\cite{Wang2024cnngenerated}, Patch-Forensics~\cite{chai2020makesfakeimagesdetectable} and Resynthesis~\cite{he2021Resynthesis}. First, we tried to reproduce the results of Abdullah et al. We could reproduce all the results except for DE-FAKE~\cite{Sha2023DE-FAKE} and  Patch-Forensics~\cite{chai2020makesfakeimagesdetectable}.  
For DE-FAKE~\cite{Sha2023DE-FAKE}, the maximum difference between Abdullah et al.~\cite{Sifat2024evolvingthreat} and our reproduced result is $64\%$ for the EfficientNet dataset. For Patch-Forensics~\cite{chai2020makesfakeimagesdetectable}, the maximum difference between Abdullah et al.~\cite{Sifat2024evolvingthreat} and our reproduced result  is $60.44\%$ for the ViT dataset. 
We attempted to clarify this discrepancy via personal communication with the authors. At the time of this submission, the discrepancy remains unresolved.
For the other three models DCT~\cite{Ricker2024detection}, CNN-F~\cite{Wang2024cnngenerated} and Resynthesis~\cite{he2021Resynthesis}, we could reproduce the result with a deviation of under $3\%$ from the result of Abdullah et al., which could be attributed to the version differences of machine learning libraries. Table~\ref{Tab:DetectorTransposed} summarizes our results. We then implemented our proposal into the five selected models. For a fair comparison, we used SyleCLIP datasets (labeled as Sec3, non-adversarial) from Abdullah et al.~\cite{Sifat2024evolvingthreat} for retraining each model following their paper.
Our results are summarized in Table~\ref{Tab:DetectorTransposed}, showing clear performance improvement for each defense after implementing our proposal. We also depict our modifications compared to the original models, except for DCT and CNN-F, as their papers lack architectural figures. 

In DCT~\cite{Ricker2024detection}, we replaced the raw log-DCT coefficients with multi-band DCT features, computing patch-wise fourth-order spectral statistics (mean, variance, skewness, kurtosis) over low, mid, and high frequencies, yielding a compact, distribution-aware representation for diffusion deepfake detection. We retrained the model using a single A100 80 GB GPU within 24 hours for the best result.
For the modified DCT, the maximum improvement of $88.9\%$ reduction in recall degradation observed in the case of CLIP-ResNet, and the minimum improvement of $51.6\%$ reduction in recall degradation observed in case of EfficientNet. 

For DE-FAKE~\cite{Sha2023DE-FAKE}, we augmented it by integrating patch-based DCT fourth-order spectral statistics (see Figure~\ref{Fig:network_architecture_de-fake}, in the appendix). Images are first decomposed into patches; after that low, mid and high frequency bands are summarized using mean, variance, skewness, and kurtosis, which are concatenated with CLIP image–text embeddings for enhanced diffusion deepfake detection. We retrained the model using a single A100 80 GB GPU within 24 hours for the best result. For the modified DE-FAKE, the maximum improvement of $86.5\%$ reduction in recall degradation is observed for EfficientNet, and the minimum improvement of $70.3\%$ reduction in recall degradation is observed for ViT.

For CNN-F~\cite{Wang2024cnngenerated}, we extended the original detector by incorporating patch-based DCT fourth-order spectral statistics. Images are first decomposed into fixed-size blocks, and their DCT coefficients are computed using mean, variance, skewness, and kurtosis, which are fused with CNN features to improve robustness against generative artifacts. The model was then retrained using a single A100 80 GB GPU, which took a total runtime of 1 day and 15 hours, and we got the best result at the 23rd epoch. For the modified CNN-F, the maximum improvement of $78.8\%$ reduction in recall degradation is observed for CLIP-ResNet, and the minimum improvement of $61.5\%$ reduction in recall degradation for EfficientNet. 

For Resynthesis~\cite{he2021Resynthesis}, we extended the detector by augmenting spatial residual features with global frequency-domain statistics derived from the DCT, capturing fourth-order distributional shifts introduced by diffusion and GAN-based generators (see Figure~\ref{Fig:network_architecture_Re-synthesis}, in the appendix).
Unlike the original model, our DCT statistics are non-learned, global descriptors extracted from reconstruction residuals and fused with Convolutional Neural Network (CNN) features. The model was then retrained for 100 epochs (following the original paper) using a single A100 80 GB GPU, which took a total runtime of 4 days and 3 hours. 
For the modified Resynthesis, the maximum improvement of $66.3\%$ reduction in recall degradation is observed for CLIP-ResNet, and the minimum improvement of $47.9\%$ reduction in recall degradation for EfficientNet.

For Patch-Forensics~\cite{chai2020makesfakeimagesdetectable}, we extended the original architecture by integrating a DCT-based statistical branch that captures fourth-order frequency statistics (mean, variance, skewness, and kurtosis), computed over the 2D DCT coefficients of the input image (see Figure~\ref{Fig:network_architecture_patch-forensics}, in the appendix).
These global frequency descriptors are then fused with Convolutional Neural Network (CNN) features, while patch shuffling is introduced for data augmentation to emphasize local artifact inconsistencies and improve generalization.  The model was than retrained for 1000 epochs (following the original paper) using a single A100 80 GB GPU, which took a total runtime of 4 days and 6 hours.
For the modified Patch-Forensics, the maximum improvement of $31.6\%$ reduction in recall degradation is observed for ViT, and the minimum improvement of $19.5\%$ reduction in recall degradation for CLIP-ResNet.

\subsection{Ablation Study}
We performed ablation studies only for $D^3$~\cite{yang2025d3}, where we integrated all of our methods.

\subhead{Effects of different disruptions} Unlike $D^3$~\cite{yang2025d3}, we applied various disruptions, e.g., patch-shuffling, random rotations (e.g., 90, 180, 270), flipping of the image at the patch level. 
Table~\ref{Tab:Disruptions} summarizes the results for integrating our techniques in $D^3$ only. We studied their effects on the model performance without introducing the third branch of content-agnostic features in the original model architecture of $D^3$. It is evident that random flip does not improve the result, while patch-shuffling alone provides the best outcome. Although no performance gain came from adding random rotation, we still kept it for robustness against the use of directionality in images to create deepfakes, following Lorch et al.~\cite{Lorch2024landscape}. We did not conduct any ablation studies after adding the third MM-BSN branch as integrating it for content-agnostic features is highly resource-intensive and increases training time very significantly.

The $D^3$ integration took approximately 14 hours/epoch on an Nvidia H100-96GB, and achieving convergence required over two weeks. A full ablation isolating the MM-BSN branch would require training four model variants (with/without MM-BSN crossed with with/without fourth-order-statistics), each taking comparable time, amounting to roughly 6–8 weeks of additional H100-time. Similarly, the five weaker detectors may take approximately 2–3 months of additional A100/H100-time. \textbf{Note:} Table~\ref{Tab:DetectorTransposed}  shows that DCT fourth-order statistics are the critical driver, yielding up to 88.9\%  recall degradation reduction (for the DCT detector~\cite{Ricker2024detection}, without MM-BSN/triple-path learning).

\begin{table}[!htb]

\setlength{\tabcolsep}{1pt}
\centering
\begin{tabularx}{0.47\textwidth}{l >{\centering\arraybackslash}X >{\centering\arraybackslash}X}
\toprule[1pt]
Defense & Acc & AP\\
\midrule[.8pt]
Patch-Shuffling & \textbf{0.819} & \textbf{0.909} \\
Patch-Shuffling + Random Rotations & 0.812 & 0.902 \\
Patch-Shuffling + Random Rotations + Flip & 0.795 & 0.881 \\
\bottomrule[1pt]
\end{tabularx}
\caption{Effects of different disruption strategies evaluated on the Surrogate\_StyleCLIP dataset~\cite{Sifat2024evolvingthreat}}
\label{Tab:Disruptions}
\end{table}

\subhead{Pooling order}
Table~\ref{Tab:pooling_order} summarizes the effects of different pooling orders on the modified $D^3$ model~\cite{yang2025d3} performance against adversarial attacks. The results show that fourth-order pooling consistently achieves the best performance.

\begin{table}[!htb]

\setlength{\tabcolsep}{3pt}
\centering
\begin{tabularx}{0.45\textwidth}{l >{\centering\arraybackslash}X >{\centering\arraybackslash}X}
\toprule[1pt]
Pooling Order & Acc & AP \\
\midrule[.8pt]
First-order statistics & 0.8530 & 0.9257 \\
Second-order statistics & 0.7900 & 0.8800 \\
Third-order statistics & 0.7960 & 0.7725 \\
Fourth-order statistics & \textbf{0.9715} & \textbf{0.9795} \\
\bottomrule[1pt]
\end{tabularx}
\caption{Performance of our defense evaluated on the Surrogate\_StyleCLIP dataset~\cite{Sifat2024evolvingthreat} w.r.t.\ different pooling orders}
\label{Tab:pooling_order}

\end{table}

\section{Discussion}

\subsection{Variations in Performance Gain}
Our proposal clearly improved all the SOTA detectors we experimented with (see Table~\ref{Tab:DetectorTransposed}). However, the magnitude of improvement varies according to the baseline's architectural priors. Our method provides the most significant gains (up to $88.9\%$ reduction in $\Delta R$) for DCT~\cite{Ricker2024detection}, as it uses mean (first-order statistics), variance and the power spectrum (second-order statistics). The attacks of Abdullah et al.~\cite{Sifat2024evolvingthreat} force the adversarial generator to prioritize specific semantic features (e.g., a smile) while satisfying the ``must look real to the surrogate'' constraint. The generator often produces textures that are ``cleaner'' or more ``standardized'' than a raw fake image, against which DCT~\cite{Ricker2024detection} is easily fooled as these semantic changes shift the first- and second-order statistics into the ``real'' range (i.e., DCT looks for specific spectral peaks or mean/variance shifts). 
Similarly,  CNN-F~\cite{Wang2024cnngenerated}, which typically relies on first- or second-order spectral statistics, shows significant improvement with our updates in the network architecture. By integrating fourth-order moments, our model identifies high-order statistical anomalies in the residual domain that are not easily masked by adversarial optimization. 

Conversely, the performance gain over Patch-Forensics~\cite{chai2020makesfakeimagesdetectable} is relatively more moderate, ranging from $19.5\%$ to $31.6\%$. We attribute this behavior to a partial overlap in inductive biases; Patch-Forensics is specifically designed to detect local textural inconsistencies by restricting the classifier's receptive field to small image patches. Our architecture adopts a similar strategy by extracting $K=16$ random patches of $224 \times 224$ resolution, processed via a ResNet50 backbone~\cite{he2016deep}. This patch-wise approach inherently avoids the global semantic shortcuts that allow adversarial attacks to fool whole-image detectors. However, our model achieves superior robustness by augmenting these spatial features with a dedicated DCT-based branch. While Patch-Forensics relies on spatial RGB distributions, our model incorporates a $20,480$-dimensional frequency feature vector derived from $8 \times 8$ and $16 \times 16$ block-wise DCT transforms. By calculating the mean, variance, skewness, and kurtosis across these blocks, our 
classifier captures high-order statistical anomalies, such as the spectral ``spikes'' and asymmetric residual distribution characteristics of GAN-based upsampling~\cite{frank2020leveraging, Wang2024cnngenerated}, which are often smoothed in the spatial domain to evade local texture classifiers. The fusion of patch-based spatial features with these multi-order frequency invariants allows our model to maintain high detection accuracy even when adversarial manipulations successfully preserve local pixel coherence.

\subsection{Possible Reasons for Our Improvements}
Even though Abdullah et al.~\cite{Sifat2024evolvingthreat} use StyleCLIP (a foundation model-based interface), the actual image's pixels are still being produced by the underlying GANs, i.e., StyleGAN2, which is their primary tool for crafting adversarial images. StyleGAN2, like most GANs, utilizes upsampling layers (e.g., transposed convolutions) that inherently introduce periodic correlations between pixels~\cite{Wang2024cnngenerated}. As demonstrated by Frank et al.~\cite{frank2020leveraging}, these structural footprints manifest as distinct spectral spikes in the frequency domain. While the adversarial attacks of Abdullah et al.~\cite{Sifat2024evolvingthreat} update the generator's weights to modify semantic content and evade CLIP-based surrogates, the structural nature of the underlying upsampling architecture remains unchanged, which may inherently introduce \emph{periodic artifacts}, in the form of  periodic pixel correlations that manifest as high-frequency spikes in the residual frequency domain~\cite{Wang2024cnngenerated, frank2020leveraging}, which we used to identify adversarial fake. %

\begin{figure*}[htb]
\centering

\begin{subfigure}[t]{0.22\textwidth}
    \centering
    \includegraphics[width=\linewidth]{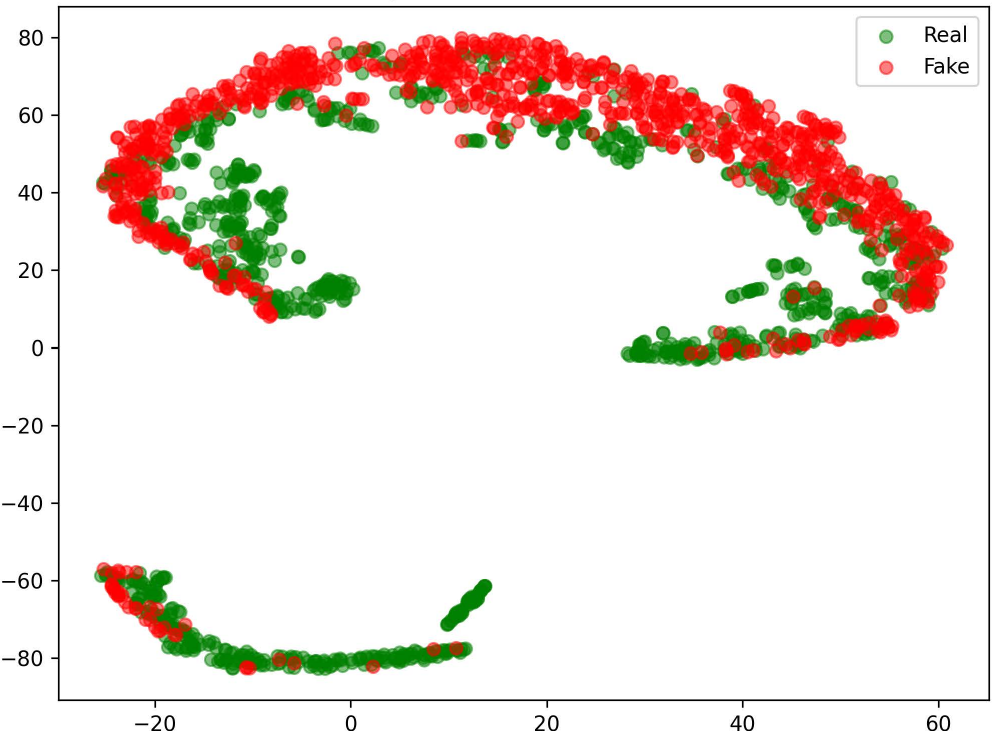}
    \caption{Original (Resynthesis)}
    \label{fig:tsne_original_rs}
\end{subfigure}
\hfill
\begin{subfigure}[t]{0.22\textwidth}
    \centering
    \includegraphics[width=\linewidth]{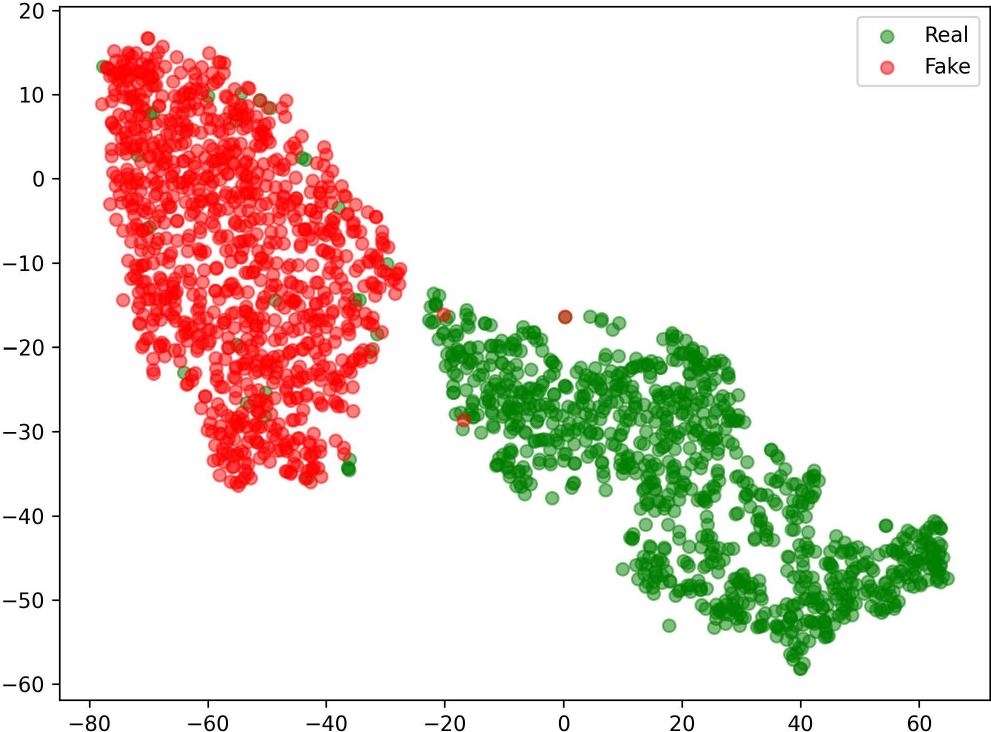}
    \caption{4th-order DCT (Resynthesis)}
    \label{fig:tsne_dct_rs}
\end{subfigure}
\hfill
\begin{subfigure}[t]{0.24\textwidth}
    \centering
    \includegraphics[width=\linewidth]{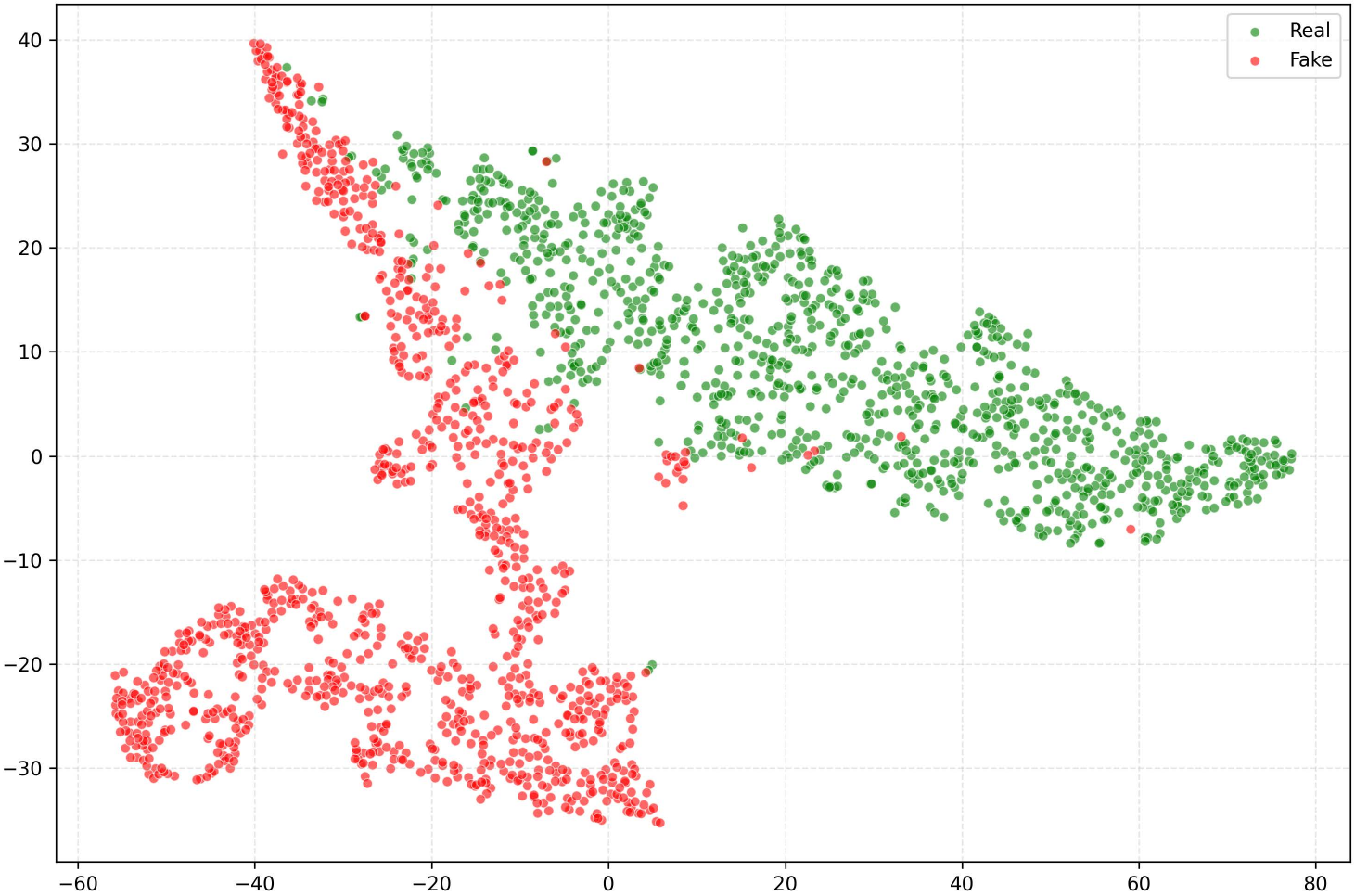}
    \caption{Original (CNN-F)}
    \label{fig:tsne_original_cnnf}
\end{subfigure}
\hfill
\begin{subfigure}[t]{0.24\textwidth}
    \centering
    \includegraphics[width=\linewidth]{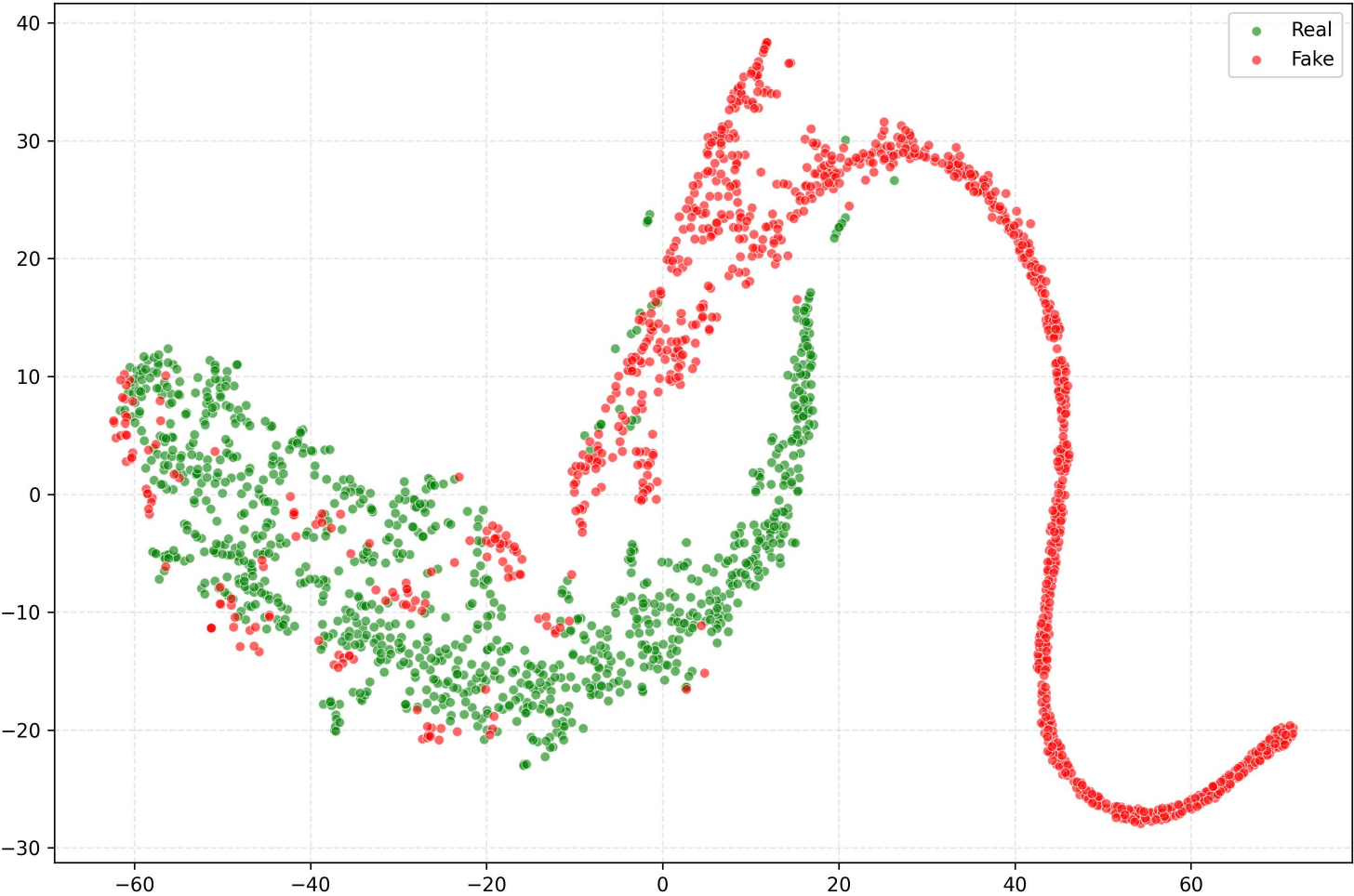}
    \caption{4th-order DCT (CNN-F)}
    \label{fig:tsne_dct_cnnf}
\end{subfigure}

\caption{t-SNE visualization of feature embeddings across two detectors. 
From left to right: original and DCT-based fourth-order representations for Resynthesis and CNN-F.} 
\label{fig:tsne_combined}

\end{figure*}

The attack by Abdullah et al.~\cite{Sifat2024evolvingthreat} is guided by surrogate classifiers (e.g., EfficientNet, ViT, or CLIP), which primarily focus on semantic features and low-to-mid frequency energy (first- and second-order statistics), leaving higher-order properties of residual-frequency distributions largely unconstrained. Our framework exploits this blind spot by (i) integrating semantic disruption (patch shuffling and rotation), which reduces reliance on global contextual cues and prevents the model from exploiting high-level semantics; (ii) content-agnostic residual features, which suppress semantic information and emphasize generator-induced noise patterns; and (iii) fourth-order frequency statistics, which capture leptokurtic residual distributions that persist under semantic manipulation. Importantly, the observed improvements are not attributable to increased feature dimensionality or model capacity. As shown in Table~\ref{Tab:pooling_order}, performance varies non-monotonically with statistical order: second- and third-order statistics do not outperform first-order features, while fourth-order statistics yield a substantial gain. This demonstrates that the improvement arises from the specific information captured by kurtosis, rather than from simply adding more features.
As established in prior frequency analyses~\cite{Wang2024cnngenerated, frank2020leveraging}, periodic artifacts from GAN upsampling layers manifest as high-magnitude outliers/spikes in the residual frequency domain, creating a distribution, which is both \emph{leptokurtic} and asymmetric~\cite{Hulzebosch2019detecting}.
Such distributions exhibit a sharper central peak and ``heavy tails,'' indicating a higher probability of extreme values or outliers compared to a normal distribution.
In our framework, these outliers correspond to the spectral spikes induced by generative upsampling layers.  
From a theoretical perspective (see Appendix~\ref{sec:theory}), the first four moments provide a minimal characterization of distributional shape, with kurtosis capturing tail heaviness and outlier prevalence. In our setting, adversarial images exhibit leptokurtic residual-frequency distributions due to structural artifacts introduced by generative models. Our model is effective because while the mean and variance may be manipulated to appear real, the skewness and kurtosis remain characteristically fake, allowing our multi-order statistical pooling to expose the generative traces that semantic-based attacks overlook (see Appendix~\ref{sec:kurtosis-avg} for differences in average kurtosis values between real and synthetic images).

\subsection{Feature Space Visualization via t-SNE}
With two of the selected detectors Resynthesis~\cite{he2021Resynthesis} and CNN-F~\cite{Wang2024cnngenerated}, we conducted a more in-depth analysis of the learned feature space to evaluate the impact of our DCT-based fourth-order statistical features on latent space representation. For this analysis, we utilized t-SNE~\cite{JMLR:v9:vandermaaten08a} to visualize the extracted features; see Figure~\ref{fig:tsne_combined}.  
For Resynthesis (Figures~\ref{fig:tsne_original_rs}-\ref{fig:tsne_dct_rs}), real and synthetic images form more separable clusters in the learned feature space in the case of DCT-based fourth-order statistics features. The t-SNE visualizations reveal that the DCT-based fourth-order statistics-enhanced model learns a more discriminative feature space, where real and synthetic images form compact and well-separated clusters. In contrast, the original baseline model (Figure~\ref{fig:tsne_original_rs}) exhibits significant overlaps between classes, consistent with its lower accuracy and recall. The DCT-based fourth-order statistics representation (Figure~\ref{fig:tsne_dct_rs}) depicts a clearer separation between real and fake samples compared to the original spatial-domain features, consistent with the improved recall and accuracy. 

We also compare the t-SNE embeddings of the baseline CNN-F against our proposed model. Figure~\ref{fig:tsne_original_cnnf} shows that the original model exhibits significant inter-class entanglement when processing the adversarial dataset; a substantial number of fake samples (red) is scattered within the real (green) cluster, indicating a failure to extract discriminative features under perturbation. This mixing directly translates to recall degradation, as the decision boundary cannot effectively isolate the manipulated samples. In contrast, Figure~\ref{fig:tsne_dct_cnnf} reveals that by integrating frequency-domain statistical priors, our model successfully pushes a large majority of the fake samples into a distinct, high-density cluster with significantly less real class leakage. There is relatively clear separation between the green and red distributions, suggesting that our approach successfully recaptures discriminative signals that the baseline model misses, effectively mitigating the degradation of recall in adversarial scenarios.

\subsection{Other Considerations}
\subhead{Evaluation against other attacks}
We further evaluated our enhanced-$D^3$ (realistic GenImage datasets) on the  RAID benchmark~\cite{eddoubi2025raiddatasettestingadversarial}, which includes a substantial proportion of abstract art images. This evaluation yielded a modest improvement, with AP increasing from 0.857 to 0.8725. It is important to note that RAID does not provide a corresponding non-adversarial training set. Evaluation on StatAttack~\cite{Hou2023adversarial} was not possible due to the unavailability of the dataset. In contrast, the attacks introduced by Abdullah et al.~\cite{Sifat2024evolvingthreat} are practically relevant: semantic/content-aware black-box evasion without additive pixel noise, specifically designed to preserve perceptual realism while defeating detectors. By comparison, noise-based perturbations (RAID/StatAttack) tend to degrade visual quality (over 3×worse), thereby posing a lesser threat.

\subhead{Robustness against common manipulations} Our training pipeline incorporates JPEG compression and Gaussian blur, each applied with probability 0.5 (Sec.~\ref{sec:impl-results}). The fourth-order DCT statistics are computed over frequency coefficients rather than raw pixels, adding inherent robustness to mild spatial perturbations (cropping/screenshotting). We tested our enhanced $D^3$ against the Surrogate-StyleCLIP-dataset with a few manipulations; the average-precision values are as follows: 0.9794 (75\% JPEG-compression), 0.9470 (80\% center-crop), 0.9830 (screen-shot-simulation); the baseline AP is 0.9795.

\subhead{Overfitting mitigation}
Our design mitigates overfitting to shuffling-induced artifacts through several mechanisms.
First, patch-shuffling is applied randomly at training time with varying patch boundaries, thereby preventing the model from learning fixed discontinuity patterns. Second, because patch-shuffling is applied equally to both real and fake images, any artifact introduced by shuffling is class-independent, i.e., non-discriminative. Third, as shown in Table~\ref{Tab:Disruptions}, patch shuffling alone yields an accuracy of 0.819 on the StyleCLIP adversarial dataset, identical to that of the original $D^3$, indicating that performance gains are not driven by spurious discriminative cues introduced by shuffling. Finally, during inference, the full unshuffled image is also processed, so the model never relies exclusively on the shuffled representation.

\subhead{Adversarial attacks against the proposed features}
We performed some preliminary experiments. By explicitly minimizing the error between the adversarial and authentic kurtosis across radial-frequency bands, we forced the optimizer to sculpt more organic deepfakes, resulting in a $\sim$30\% recall degradation  (against our improved DCT detector). This needs more experiments for validation.

\section{Conclusion}
By evaluating seven recent state-of-the-art detectors, we reconfirm the serious performance degradation reported by Abdullah et al.~\cite{Sifat2024evolvingthreat} under adversarial attacks. 
Our proposed framework is designed to enhance the robustness of existing state-of-the-art detectors against such known adversarial attacks.
Our results demonstrate that, when trained on datasets with sufficient semantic diversity and high visual fidelity, the proposed modifications can substantially improve detector robustness without requiring adversarial training data. In particular, incorporating diverse semantic content within training datasets helps mitigate reliance on spurious semantic artifacts, while ensuring that each class includes samples generated by a wide range of high-quality generative models.
Furthermore, our findings highlight the importance of emphasizing content-agnostic representations and higher-order statistical features in detector design. 
While our work provides a principled approach toward this direction, it also reveals several open challenges that warrant further investigation.

\section*{Acknowledgments}
We sincerely thank the anonymous Usenix Security 2026 reviewers
for their insightful suggestions and comments. For computing and GPU support, we are grateful to David Lie (University of Toronto), and the high-performance computing facility (\emph{Speed}) at Concordia University. This research is funded by an NSERC-CSE Research Communities Grant. 
Researchers funded through the NSERC-CSE Research Communities Grants do not represent the Communications Security Establishment Canada or the Government of Canada. Any research, opinions or positions they produce as part of this initiative do not represent the official views of the Government of Canada. 
 
\section*{Ethical Considerations}

We begin our ethical analysis by identifying the primary stakeholders impacted by our work. They include: (i) victims of deepfake abuse (e.g., individuals targeted by harassment, blackmail, non-consensual intimate image abuse, or political disinformation); (ii) society at large (general public affected by misinformation and erosion of trust in digital media); (iii) developers and users of deepfake detection systems (e.g., media sharing platforms, journalists, and forensic analysts); (iv) generative AI companies and content moderation teams; (v) malicious actors who create adversarial deepfakes; and (vi) the research community working on the problem of deepfake detection.

Our research was guided by core principles for ethical computing research~\cite{kohno2023ethical}.
The ethical principles we considered are: (i) Beneficence: Our work seeks to improve the robustness of SOTA deepfake detectors against content-manipulating adversarial attacks without requiring adversarial training, thereby reducing the real-world harm caused by high-quality synthetic images used for disinformation, harassment, and blackmail; (ii) Respect for Persons: No human subjects, private data, or deceptive experiments were involved. We used only publicly available datasets; (iii) Justice: We aim for fair detection performance across diverse content types. However, we acknowledge that detection systems may still inherit biases from training data or model architectures, potentially leading to uneven performance across demographics or image styles.
We did not measure or mitigate demographic bias as the GenImage and StyleCLIP datasets we used are not annotated with demographic attributes, which will require substantial additional data collection and labeling effort beyond the scope of this work. GenImage uses ImageNet object/animal classes, and StyleCLIP is a face-only dataset with no demographic labels such as skin tone, ethnicity, or identity. Future work should audit deepfake detectors using demographically stratified benchmarks, particularly given evidence in face-recognition literature that GAN artifacts can correlate with skin tone due to training data biases. We view this as an important open problem; and (iv) Respect for Law and Public Interest: This research supports societal trust in digital media and does not involve unauthorized system access or terms-of-service violations.

The harms we identified fall into two categories.
Tangible harms include: (i) Dual-use risk: Our defense techniques could be analyzed by attackers to develop more sophisticated evasion methods; (ii) Over-reliance: Users may place excessive trust in automated detectors, leading to insufficient manual verification of critical content; and (iii) Resource disparity: Our approach requires significant GPU resources for training, which may limit adoption by under-resourced researchers or organizations.

Regarding rights-based harms, no direct violations of privacy, consent, or autonomy occurred, as all experiments used public datasets and no human participants.

We implemented several measures to address these risks. Our key mitigations are: (i) We exclusively used publicly available, non-sensitive datasets (GenImage and StyleCLIP), and pretrained models; (ii) We open-source our code and improved models to promote transparency, reproducibility, and equitable access; (iii) We explicitly avoided low-quality adversarial datasets (e.g., RAID) to prevent misleading evaluations; (iv) We did not perform adversarial training, reducing the risk of exposing harmful content during training; and (v) All attacks we tested on new detectors are already public~\cite{Sifat2024evolvingthreat}. Our attack experiments were performed locally, in a controlled lab setting within our own lab machines, or in our school's computing cluster. We did not experiment with live systems or services. We also do not propose any new attack methodologies for deepfake image generation, or bypassing the current SOTA detectors. Our techniques are designed to help defenders to improve robustness of their deepfake image detectors.

Despite these efforts, some challenges remain. The unmitigated/ residual risks include: (i) An ongoing adversarial arms race may continue, as attackers can adapt to our improvements; and (ii) Generalization may be limited in certain cultural or demographic contexts.

After weighing these factors, we decided to conduct and publish this research because the potential benefits (significantly stronger defenses against current adversarial deepfake attacks) substantially outweigh the associated risks. Although publishing robustness evaluations and defense strategies may indirectly help adversaries understand current detector limitations, the attacks evaluated in this work are already publicly known~\cite{Sifat2024evolvingthreat}, and our work does not introduce new attack methodologies or provide operational guidance for misuse. Instead, the primary contribution is improving defensive robustness against existing threats. Prior work has documented broad societal risks of deepfake technologies, including blackmail, harassment, erosion of trust in digital media, and large-scale disinformation~\cite{alanazi2024examining,veerasamy2022rising}. Legal and ethical analyses further highlight concerns related to privacy violations and identity misuse, underscoring the need for effective regulatory responses~\cite{meskys2020regulating}. These findings are consistent with reported real-world cases of deepfake misuse, involving blackmail with AI-generated intimate imagery~\cite{alanazi2024examining}, psychological harm~\cite{alanazi2025unmasking}, and political misinformation campaigns~\cite{veerasamy2022rising}. In light of these findings, we believe that openly advancing and evaluating defensive techniques serves the public interest and supports more trustworthy deployment of deepfake detection systems.
Finally, we confirm our legal compliance. No IRB approval was required, as this work does not involve human subjects or any collection/use of personal information. All datasets are publicly available for research use. We did not test on live systems, violate terms of service, or engage in unauthorized access.

\section*{Open Science}
Code and datasets are available at: \url{https://doi.org/10.5281/zenodo.20327568}. 
We release the six detectors integrated in our proposed framework. Detailed instructions for building and running the code are included in the \texttt{README.md} file.

The size of our training subset (36,000 images) is 13.2 GB, which exceeds the storage limitations of the repository platform. However, we provide sample images sufficient for running the code, though not for full reproducibility of our results. The complete dataset can be made available upon request (e.g., via Google Drive or other file-sharing services).

\bibliographystyle{plainurl}

\bibliography{references}

@article{Nightingale2022AISynthesized,
  author       = {Sophie J. Nightingale and Hany Farid},
  title        = {{AI-synthesized faces are indistinguishable from real faces and more trustworthy}},
  journal      = {Proceedings of the National Academy of Sciences of the United States of America (PNAS)},
  volume       = {119},
  number       = {8},
  year         = {2022}
}

@inproceedings{Tan2024upsampling,
  title={Rethinking the Up-Sampling Operations in {CNN}-based Generative Network for Generalizable Deepfake Detection},
  author={C. Tan and H. Liu and Y. Zhao and S. Wei and G. Gu and P. Liu and Y. Wei},
  booktitle={Proc. IEEE Conference on Computer Vision and Pattern Recognition},
  pages={},
  year={2024}
}

@inproceedings{Sarkar2024shadows,
  title={Shadows Don’t Lie and Lines Can’t Bend! {G}enerative Models Don’t Know Projective Geometry...for Now},
  author={A. Sarkar and H. Mai and A. Mahapatra and S. Lazebnik and D. A. Forsyth and A. Bhattad},
  booktitle={Proc. IEEE Conference on Computer Vision and Pattern Recognition},
  pages={},
  year={2024}
}

@inproceedings{Corvi2023intriguing,
  title={Intriguing properties of synthetic images: from generative adversarial networks to diffusion models},
  author={R.Corvi and D.Cozzolino and G.Poggi and K.Nagano and L.Verdoliva},
  booktitle={Proc. IEEE Conference on Computer Vision and Pattern Recognition Workshops},
  pages={},
  year={2023}
}

@inproceedings{Lorch2024landscape,
  title={Landscape More Secure Than Portrait? {Z}ooming Into the Directionality of Digital Images With Security Implications},
  author={B. Lorch and R. Böhme},
  booktitle={Proc. 33rd USENIX Security Symposium},
  pages={6903--6920},
  year={2024}
}

@inproceedings{Podell2024sdxl,
  title={{SDXL}: Improving Latent Diffusion Models for High-Resolution Image Synthesis},
  author={D. Podell and Z. English and K. Lacey and A. Blattmann and T. Dockhorn and J. Müller and J. Penna and R. Rombach},
  booktitle={International Conference on Learning Representations},
  year={2024}
}

@inproceedings{Ojha2023ufd,
  title={Towards Universal Fake Image Detectors that Generalize Across Generative Models},
  author={Utkarsh Ojha and Yuheng Li and Yong Jae Lee},
  booktitle={Proc. IEEE Conference on Computer Vision and Pattern Recognition},
  pages={},
  year={2023}
}

@inproceedings{liu2020globaltextureenhancementfake,
  title={Global Texture Enhancement for Fake Face Detection In the Wild},
  author={Zhengzhe Liu and Xiaojuan Qi and Philip Torr},
  booktitle={Proc. IEEE Conference on Computer Vision and Pattern Recognition},
  pages={8060--8069},
  year={2020}
}

@inproceedings{Wang2023dire,
  title={{DIRE} for Diffusion-Generated Image Detection},
  author={Z. Wang and J. Bao and W. Zhou and W. Wang and H. Hu and H. Chen and H. Li},
  booktitle={Proc. IEEE Conference on Computer Vision and Pattern Recognition},
  pages={22445--22455},
  year={2023}
}

@inproceedings{Zhang2023mmbsn,
  title={{MM-BSN}: Self-Supervised Image Denoising for Real-World with Multi-Mask},
  author={Dan Zhang and Fangfang Zhou and Yuwen Jiang and Zhengming Fu},
  booktitle={Proc. IEEE Conference on Computer Vision and Pattern Recognition Workshops},
  pages={4189--4198},
  year={2023}
}

@inproceedings{zhang2025diffusion4k,
    title={{Diffusion-4K}: Ultra-High-Resolution Image Synthesis with Latent Diffusion Models},
    author={Zhang, Jinjin and Huang, Qiuyu and Liu, Junjie and Guo, Xiefan and Huang, Di},
    year={2025},
    booktitle={Proc. IEEE Conference on Computer Vision and Pattern Recognition},
}

@inproceedings{Wang2022Blind2Unblind,
    title={{Blind2Unblind}: Self-Supervised Image Denoising with Visible Blind Spots},
    author={Zejin Wang and Jiazheng Liu and Guoqing Li and Hua Han},
    pages={2027--2036},
    year={2022},
    booktitle={Proc. IEEE Conference on Computer Vision and Pattern Recognition},
}

@inproceedings{Chu2025fire,
    title={{FIRE}: Robust Detection of Diffusion-Generated Images via Frequency-Guided Reconstruction Error},
    author={Beilin Chu and Xuan Xu and Xin Wang and Yufei Zhang and Weike You and Linna Zhou},
    pages={},
    year={2025},
    booktitle={Proc. IEEE Conference on Computer Vision and Pattern Recognition},
}

@inproceedings{Cao2025temporal,
    title={Temporal Score Analysis for Understanding and Correcting Diffusion Artifacts},
    author={Yu Cao and Zengqun Zhao and Ioannis Patras and Shaogang Gong},
    pages={},
    year={2025},
    booktitle={Proc. IEEE Conference on Computer Vision and Pattern Recognition},
}

@inproceedings{Karageorgiou2025anyresolution,
    title={Any-Resolution {AI}-Generated Image Detection by Spectral Learning},
    author={Dimitrios Karageorgiou and Symeon Papadopoulos and Ioannis Kompatsiaris and Efstratios Gavves},
    pages={18706--18717},
    year={2025},
    booktitle={Proc. IEEE Conference on Computer Vision and Pattern Recognition},
}

@inproceedings{Huang2025sida,
    title={{SIDA}: Social Media Image Deepfake Detection, Localization and Explanation with Large Multimodal Model},
    author={Zhenglin Huang and Jinwei Hu and Xiangtai Li and Yiwei He and
Xingyu Zhao and Bei Peng and Baoyuan Wu and Xiaowei Huang and Guangliang Cheng},
    pages={28831--28841},
    year={2025},
    booktitle={Proc. IEEE Conference on Computer Vision and Pattern Recognition},
}

@inproceedings{yang2025d3,
  title={{$D^3$}: Scaling Up Deepfake Detection by Learning from Discrepancy},
  author={Yang, Yongqi and Qian, Zhihao and Zhu, Ye and Russakovsky, Olga and Wu, Yu},
  booktitle={Proc. IEEE Conference on Computer Vision and Pattern Recognition\balance},
  year={2025}
}

@inproceedings{Cheng2025cospy,
  title={{CO-SPY}: Combining Semantic and Pixel Features to Detect Synthetic Images by {AI}},
  author={Siyuan Cheng and Lingjuan Lyu and Zhenting Wang and Xiangyu Zhang and Vikash Sehwag},
  pages={13455--13465},
  booktitle={Proc. IEEE Conference on Computer Vision and Pattern Recognition},
  year={2025}
}

@inproceedings{Brokman2025manifold,
  title={Manifold induced biases for Zero-Shot and Few-Shot detection of generated images},
  author={Jonathan Brokman and Amit Giloni and Omer Hofman and Roman Vainshtein and Hisashi Kojima and Guy Gilboa},
  pages={1--26},
  booktitle={International Conference on Learning Representations},
  year={2025}
}

@inproceedings{Zheng2024semanticartifacts,
  title={Breaking Semantic Artifacts for Generalized {AI}-generated Image Detection},
  author={Chende Zheng and Chenhao Lin and Zhengyu Zhao and Hang Wang and Xu Guo and Shuai Liu and Chao Shen},
  booktitle={Advances in Neural Information Processing Systems},
  year={2024}
}

@inproceedings{Zhu2024genimage,
  title={{GenImage}: A Million-Scale Benchmark for Detecting {AI}-Generated Image},
  author={Mingjian Zhu and Hanting Chen and Qiangyu Yan and Xudong Huang and Guanyu Lin and Wei Li and Zhijun Tu and Hailin Hu and Jie Hu and Yunhe Wang},
  booktitle={Advances in Neural Information Processing Systems},
  pages={},
  year={2024}
}

@inproceedings{Ricker2024detection,
  title={Towards the Detection of Diffusion Model Deepfakes},
  author={Jonas Ricker and Simon Damm and Thorsten Holz and Asja Fischer},
  booktitle={Proc. Conference on Computer Vision, Imaging and Computer Graphics Theory and Applications},
  year={2024}
}

@inproceedings{Li2024masksim,
    title={{Masksim}: Detection of synthetic images by masked spectrum similarity analysis},
    author={Yanhao Li and Quentin Bammey and Marina Gardella and Tina Nikoukhah and Jean-Michel Morel and Miguel Colom and Rafael Grompone Von Gioi},
    pages={},
    year={2024},
    booktitle={Proc. IEEE Conference on Computer Vision and Pattern Recognition Workshops},
}

@inproceedings{Hou2023adversarial,
    title={Evading DeepFake Detectors via Adversarial Statistical Consistency},
    author={Yang Hou and Qing Guo and Yihao Huang and Xiaofei Xie and Lei Ma and Jianjun Zhao},
    pages={12271--12280},
    year={2023},
    booktitle={Proc. IEEE Conference on Computer Vision and Pattern Recognition},
}

@inproceedings{Wang2023dynamicgraphlearning,
    title={Dynamic graph learning with content-guided spatial-frequency relation reasoning for deepfake detection},
    author={Yuan Wang and Kun Yu and Chen Chen and Xiyuan Hu and Silong Peng},
    pages={7278--7287},
    year={2023},
    booktitle={Proc. IEEE Conference on Computer Vision and Pattern Recognition},
}

@inproceedings{Wang2024cnngenerated,
  title={{CNN}-generated images are surprisingly easy to spot...for now},
  author={Wang, Sheng-Yu and Wang, Oliver and Zhang, Richard and Owens, Andrew and Efros, Alexei A},
  booktitle={Proc. IEEE Conference on Computer Vision and Pattern Recognition},
  year={2020}
}

@misc{eddoubi2025raiddatasettestingadversarial,
      title={{RAID}: A Dataset for Testing the Adversarial Robustness of {AI}-Generated Image Detectors}, 
      author={Hicham Eddoubi and Jonas Ricker and Federico Cocchi and Lorenzo Baraldi and Angelo Sotgiu and Maura Pintor and Marcella Cornia and Lorenzo Baraldi and Asja Fischer and Rita Cucchiara and Battista Biggio},      
      note={arXiv preprint (June 9, 2025), \url{https://arxiv.org/abs/2506.03988}}
}

@misc{yu2015lsun,
  title={{LSUN}: Construction of a large-scale image dataset using deep learning with humans in the loop},
  author={Yu, Fisher and Seff, Ari and Zhang, Yinda and Song, Shuran and Funkhouser, Thomas and Xiao, Jianxiong},
  note={arXiv preprint (Jun 4, 2016), \url{https://arxiv.org/abs/1506.03365}}
}

@inproceedings{Sifat2024evolvingthreat,
    title={An Analysis of Recent Advances in Deepfake Image Detection in an
Evolving Threat Landscape},
    author={Sifat Muhammad Abdullah and Aravind Cheruvu and Shravya Kanchi and Taejoong Chung and Peng Gao and Murtuza Jadliwala and Bimal Viswanath},
    pages={91--109},
    year={2024},
    booktitle={IEEE Symposium on Security and Privacy (SP)},
}

@inproceedings{Esser2024rectifiedflow,
    title = {Scaling rectified flow transformers for high-resolution image synthesis},
    author = {Esser, Patrick and Kulal, Sumith and Blattmann, Andreas and Entezari, Rahim and M\"{u}ller, Jonas and Saini, Harry and Levi, Yam and Lorenz, Dominik and Sauer, Axel and Boesel, Frederic and Podell, Dustin and Dockhorn, Tim and English, Zion and Rombach, Robin},
    booktitle = {Proc. International Conference on Machine Learning},
    numpages = {12606 - 12633},
    year = {2024}
}

@inproceedings{Liu2023rectifiedflow,
    title = {Flow Straight and Fast: Learning to Generate and Transfer Data with Rectified Flow},
    author = {Liu, Xingchao and Gong, Chengyue and Liu, Qiang},
    booktitle = {International Conference on Learning Representations},
    year = {2023}
}

@InProceedings{sohl2015Nonequilibriumthermodynamics,
    title = 	 {Deep Unsupervised Learning using Nonequilibrium Thermodynamics},
    author = 	 {Sohl-Dickstein, Jascha and Weiss, Eric and Maheswaranathan, Niru and Ganguli, Surya},
    booktitle = 	 {Proc. International Conference on Machine Learning},
    pages = 	 {2256--2265},
    year = 	 {2015}
}

@InProceedings{Radford2021Naturallanguage,
    title = 	 {Learning Transferable Visual Models From Natural Language Supervision},
    author = 	 {Alec Radford and Jong Wook Kim and Chris Hallacy and Aditya
Ramesh and Gabriel Goh and Sandhini Agarwal and Girish Sastry and Amanda Askell and Pamela Mishkin and Jack Clark and Gretchen Krueger and Ilya Sutskever},
    booktitle = 	 {Proc. International Conference on Machine Learning},
    pages = 	 {},
    year = 	 {2021}
}

@article{Wolter2022wavelet,
  title={Wavelet-Packets for Deepfake Image Analysis and Detection},
  author={M. Wolter and F. Blanke and R. Heese and J.Garcke},
  journal={Machine Learning},
  volume={},
  pages={4295--4327},
  year={2022}
}

@article{Tran2025diffcor,
  title={{DiffCoR}: Exposing {AI-Generated} Image by Using Stable Diffusion Model Based on Consistent Representation Learning},
  author={V. -N. Tran and P. Choi and H. -S. Le and S. -H. Lee and K. -R. Kwon},
  journal={IEEE Open Journal of the Computer Society},
  volume={},
  pages={},
  year={2025}
}

@inproceedings{Jiameng2020noisescope,
    title = {{NoiseScope}: Detecting Deepfake Images in a Blind Setting},
    author = {Jiameng Pu and Neal Mangaokar and Bolun Wang and Chandan K. Reddy and Bimal Viswanath},
    booktitle = {Proc. Annual Computer Security Applications Conference},
    year = {2020}
}

@inproceedings{Sha2023DE-FAKE,
author = {Sha, Zeyang and Li, Zheng and Yu, Ning and Zhang, Yang},
title = {{DE-FAKE}: Detection and Attribution of Fake Images Generated by Text-to-Image Generation Models},
year = {2023},
booktitle = {Proc. ACM SIGSAC Conference on Computer and Communications Security},
pages = {3418–3432},
numpages = {15},
}

@inproceedings{he2016deep,
  title={Deep Residual Learning for Image Recognition},
  author={He, Kaiming and Zhang, Xiangyu and Ren, Shaoqing and Sun, Jian},
  booktitle={Proc. IEEE Conference on Computer Vision and Pattern Recognition},
  year={2016}
}

@article{zhang2017beyond,
  title={Beyond a Gaussian Denoiser: Residual Learning of Deep {CNN} for Image Denoising},
  author={Zhang, Kai and Zuo, Wangmeng and Chen, Yunjin and Meng, Deyu and Zhang, Lei},
  journal={IEEE Transactions on Image Processing},
  year={2017}
}

@inproceedings{kid-test,
      title={Demystifying {MMD GANs}}, 
      author={Miko\l{}aj Bi\'{n}kowski and Danica J. Sutherland and Michael Arbel and Arthur Gretton},
      booktitle = {International Conference on Learning Representations},
      year={2018}
}

@article{JMLR:v9:vandermaaten08a,
  author  = {Laurens van der Maaten and Geoffrey Hinton},
  title   = {Visualizing Data using {t-SNE}},
  journal = {Journal of Machine Learning Research},
  year    = {2008},
  volume  = {9},
  number  = {86},
  pages   = {2579--2605},
}

@inproceedings{Sun2019HRNet,
  author    = {Sun, Ke and Xiao, Bin and Liu, Dong and Wang, Jingdong},
  title     = {Deep High-Resolution Representation Learning for Human Pose Estimation},
  booktitle = {Proc. IEEE Conference on Computer Vision and Pattern Recognition},
  year      = {2019},
  pages     = {569--578}
}

@inproceedings{Chen2020SimCLR,
  author    = {Chen, Ting and Kornblith, Simon and Norouzi, Mohammad and Hinton, Geoffrey},
  title     = {A Simple Framework for Contrastive Learning of Visual Representations},
  booktitle = {Proc. International Conference on Machine Learning},
  year      = {2020},
  pages     = {}
}

@INPROCEEDINGS{MesoNet2018afchar,
  author={Afchar, Darius and Nozick, Vincent and Yamagishi, Junichi and Echizen, Isao},
  booktitle={IEEE International Workshop on Information Forensics and Security}, 
  title={MesoNet: a Compact Facial Video Forgery Detection Network}, 
  year={2018},
  pages={1-7},
  }

@inproceedings{Kingma2014AdamAM,
    title = {Adam: A Method for Stochastic Optimization},
    author = {Diederik P. Kingma and Jimmy Ba},
    booktitle = {Proc. International Conference on Learning Representations},
    year = {2015}
}

@inproceedings{frank2020leveraging,
  title={Leveraging frequency analysis for deep fake image recognition},
  author={Frank, Joel and Eisenhofer, Thorsten and Sch{\"o}nherr, Lea and Fischer, Asja and Kolossa, Dorothea and Holz, Thorsten},
  booktitle={Proc. International Conference on Machine Learning},
  pages={3247--3258},
  year={2020}
}

@InProceedings{pmlr-v97-tan19a,
  title = 	 {{E}fficient{N}et: Rethinking Model Scaling for Convolutional Neural Networks},
  author =       {Tan, Mingxing and Le, Quoc},
  booktitle = 	 {Proc. International Conference on Machine Learning},
  pages = 	 {6105--6114},
  year = 	 {2019},
  volume = 	 {97},
  publisher =    {PMLR},
}

@InProceedings{dosovitskiy2021imageworth16x16words,
      title={An Image is Worth 16x16 Words: Transformers for Image Recognition at Scale}, 
      author={Alexey Dosovitskiy and Lucas Beyer and Alexander Kolesnikov and Dirk Weissenborn and Xiaohua Zhai and Thomas Unterthiner and Mostafa Dehghani and Matthias Minderer and Georg Heigold and Sylvain Gelly and Jakob Uszkoreit and Neil Houlsby},
      year={2021},
      booktitle={International Conference on Learning Representations}, 
}

@inproceedings{
hu2022lora,
title={Lo{RA}: Low-Rank Adaptation of Large Language Models},
author={Edward J Hu and Yelong Shen and Phillip Wallis and Zeyuan Allen-Zhu and Yuanzhi Li and Shean Wang and Lu Wang and Weizhu Chen},
booktitle={International Conference on Learning Representations},
year={2022}
}

@inproceedings{li2020rethinking,
  title={Rethinking the Hyperparameters for Fine-tuning},
  author={Li, Hao and Chaudhari, Pratik and Yang, Hao and Lam, Michael and Ravichandran, Avinash and Bhotika, Rahul and Soatto, Stefano},
  booktitle={International Conference on Learning Representations},
  year={2020},
}

@inproceedings{Hulzebosch2019detecting,
  title={Detecting {CNN}-generated facial images in real-world scenarios},
  author={Nils Hulzebosch and Sarah Ibrahimi and Marcel Worring},
  booktitle={Proc. IEEE Conference on Computer Vision and Pattern Recognition  Workshops},
  year={2020}
}

@article{everingham2010pascal,
  title={The {PASCAL} visual object classes ({VOC}) challenge},
  author={Everingham, Mark and Van Gool, Luc and Williams, Christopher KI and Winn, John and Zisserman, Andrew},
  journal={International Journal of Computer Vision},
  volume={88},
  number={2},
  pages={},
  year={2010}
}

@inproceedings{schuhmann2021laion400m,
  title={{LAION-400M}: Open Dataset of {CLIP}-Filtered 400 Million Image-Text Pairs},
  author={Schuhmann, Christoph and Vencu, Richard and Beaumont, Romain and Kaczmarczyk, Robert and Mullis, Clayton and Katta, Aarush and Coombes, Theo and Jitsev, Jenia and Komatsuzaki, Aran},
  booktitle = {NeurIPS Workshop on Machine Learning for Creativity and Design},
  year={2021}
}

@article{deng2009imagenet,
  title={{ImageNet}: A Large-Scale Hierarchical Image Database},
  author={Deng, Jia and Dong, Wei and Socher, Richard and Li, Li-Jia and Li, Kai and Fei-Fei, Li},
  journal={Proc. IEEE Conference on Computer Vision and Pattern Recognition},
  year={2009}
}

@inproceedings{karras2018progressive,
  title={Progressive Growing of {GANs} for Improved Quality, Stability, and Variation},
  author={Karras, Tero and Aila, Timo and Laine, Samuli and Lehtinen, Jaakko},
  booktitle={International Conference on Learning Representations},
  year={2018}
}

@inproceedings{brock2019biggan,
  title={Large Scale {GAN} Training for High Fidelity Natural Image Synthesis},
  author={Brock, Andrew and Donahue, Jeff and Simonyan, Karen},
  booktitle={International Conference on Learning Representations},
  year={2019}
}

@inproceedings{karras2019stylegan,
  title={A Style-Based Generator Architecture for Generative Adversarial Networks},
  author={Karras, Tero and Laine, Samuli and Aila, Timo},
  booktitle={Proc. IEEE Conference on Computer Vision and Pattern Recognition},
  year={2019}
}

@inproceedings{ramesh2021zeroshot,
  author = {Ramesh, Aditya and Pavlov, Mikhail and Goh, Gabriel and Gray, Scott and Voss, Chelsea and Radford, Alec and Chen, Mark and Sutskever, Ilya},
  title = {Zero-Shot Text-to-Image Generation},
  booktitle = {Proc. International Conference on Machine Learning },
  year = {2021},
  pages = {8821--8831}
}

@inproceedings{liu2022generatedimages,
  title={Detecting Generated Images by Real Images},
  author={Bo Liu and Fan Yang and Xiuli Bi and Bin Xiao and Weisheng Li and Xinbo Gao},
  booktitle={Proc. European Conference on Computer Vision },
  year={2022}
}

@inproceedings{he2021Resynthesis,
      title={Beyond the Spectrum: Detecting Deepfakes via Re-Synthesis}, 
      author={Yang He and Ning Yu and Margret Keuper and Mario Fritz},
      year={2021},
      booktitle={Proc. International Joint Conference on Artificial Intelligence},
}

@inproceedings{chai2020makesfakeimagesdetectable,
      title={What makes fake images detectable? {U}nderstanding properties that generalize}, 
      author={Lucy Chai and David Bau and Ser-Nam Lim and Phillip Isola},
      year={2020},
      booktitle={Proc. European Conference on Computer Vision},
}

@inproceedings{rombach2022ldm,
  title={High-Resolution Image Synthesis with Latent Diffusion Models},
  author={Rombach, Robin and Blattmann, Andreas and Lorenz, Dominik and Esser, Patrick and Ommer, Bj{\"o}rn},
  booktitle={Proc. IEEE Conference on Computer Vision and Pattern Recognition},
  year={2022}
}

@inproceedings{nichol2022glide,
  title={{GLIDE}: Towards Photorealistic Image Generation and Editing with Text-Guided Diffusion Models},
  author={Nichol, Alexander Quinn and Dhariwal, Prafulla and Ramesh, Aditya and Shyam, Pranav and Mishkin, Pamela and McGrew, Bob and Sutskever, Ilya and Chen, Mark},
  booktitle={Proc. International Conference on Machine Learning},
  year={2022}
}

@inproceedings{gu2022vqdm,
  author = {Gu, Shuyang and Bao, Jianmin and Chen, Dong and Wen, Fang},
  title = {Vector Quantized Diffusion Model for Text-to-Image Synthesis},
  booktitle = {Proc. IEEE Conference on Computer Vision and Pattern Recognition},
  year = {2022},
  pages = {10696--10706}
}

@inproceedings{gu2022wukong,
  title     = {Wukong: A 100 Million Large-scale Chinese Cross-modal Pre-training Benchmark},
  author    = {Gu, Jiaxi and Meng, Xiaojun and Lu, Guansong and Hou, Lu and Niu, Minzhe and Liang, Xiaodan and Yao, Lewei and Huang, Runhui and Zhang, Wei and Jiang, Xin and Xu, Chunjing and Xu, Hang},
  booktitle = {Advances in Neural Information Processing Systems (NeurIPS'22)},
  year      = {2022}
}

@inproceedings{dhariwal2021adm,
  author = {Dhariwal, Prafulla and Nichol, Alex},
  title = {Diffusion Models Beat {GANs} on Image Synthesis},
  booktitle = {Advances in Neural Information Processing Systems},
  year = {2021}
}

@misc{midjourney2022,
  author       = {{Midjourney, Inc.}},
  title        = {{Midjourney}: Proprietary Text-to-Image Generation System},
  howpublished = {\url{https://www.midjourney.com/}},
  year         = {2022},
  note         = {Online; accessed Oct. 23, 2025}
}

@inproceedings{vladimir-etal-2024-kandinsky,
    title = "Kandinsky 3: Text-to-Image Synthesis for Multifunctional Generative Framework",
    author = "Vladimir, Arkhipkin  and
      Vasilev, Viacheslav  and
      Filatov, Andrei  and
      Pavlov, Igor  and
      Agafonova, Julia  and
      Gerasimenko, Nikolai  and
      Averchenkova, Anna  and
      Mironova, Evelina  and
      Anton, Bukashkin  and
      Kulikov, Konstantin  and
      Kuznetsov, Andrey  and
      Dimitrov, Denis",
    booktitle = "Proc. Conference on Empirical Methods in Natural Language Processing: System Demonstrations",
    year = "2024",
    pages = "475--485",
}

@misc{toi_ai_blackmail_suicide_2025,
  author       = {{IndiaTimes.com}},
  title        = {{Faridabad AI shocker: College student dies by suicide; friend made obscene deepfakes, blackmailed him}},
  note = {News article (Oct 27, 2025), \url{https://timesofindia.indiatimes.com/city/gurgaon/blackmailed-by-ai-generated-pictures-student-dies-by-suicide-in-faridabad/articleshow/124833566.cms}. Accessed: Feb 4, 2026}
}

@misc{guardian_deepfake_porn_schools_2025,
  author       = {{TheGuardian.com}},
  title        = {The rise of deepfake pornography in schools},
  note         = {News article (Dec 2, 2025), \url{https://www.theguardian.com/society/ng-interactive/2025/dec/02/the-rise-of-deepfake-pornography-in-schools}. Accessed: Feb 4, 2026}
}

@misc{guardian2026maduroai,
  author       = {{TheGuardian.com}},
  title        = {{AI} images of {Maduro} capture reap millions of views on social media},
  note         = {News article (Jan 6, 2026), \url{https://www.theguardian.com/technology/2026/jan/05/maduro-venezuela-ai-images}. Accessed: Feb 4, 2026}
}

@inproceedings{kohno2023ethical,
  title={Ethical frameworks and computer security trolley problems: Foundations for conversations},
  author={Kohno, Tadayoshi and Acar, Yasemin and Loh, Wulf},
  booktitle={32nd USENIX Security Symposium},
  pages={5145--5162},
  year={2023}
}

@article{alanazi2025unmasking,
  title={Unmasking deepfakes: a multidisciplinary examination of social impacts and regulatory responses},
  author={Alanazi, Sami and Asif, Seemal and Caird-Daley, Antoinette and Moulitsas, Irene},
  journal={Human-Intelligent Systems Integration},
  pages={131--153},
  year={2025}
}

@article{alanazi2024examining,
  title={Examining the societal impact and legislative requirements of deepfake technology: A comprehensive study},
  author={Alanazi, Sami and Asif, Seemal and Moulitsas, Irene},
  journal={International Journal of Social Science and Humanity},
  year={2024}
}

@article{meskys2020regulating,
  title={Regulating deep fakes: legal and ethical considerations},
  author={Meskys, Edvinas and Kalpokiene, Julija and Jurcys, Paul and Liaudanskas, Aidas},
  journal={Journal of Intellectual Property Law \& Practice},
  volume={15},
  number={1},
  pages={24--31},
  year={2020}
}

@inproceedings{veerasamy2022rising,
  title={Rising above misinformation and deepfakes},
  author={Veerasamy, Namosha and Pieterse, Heloise},
  booktitle={International Conference on Cyber Warfare and Security},
  volume={17},
  number={1},
  pages={340--348},
  year={2022}
}

\appendix

\section*{Appendix}
\section{Theoretical Justification For First–Fourth Moments}
\label{sec:theory}
The following theorem motivates first–fourth moments as the minimal set capturing location/scale plus the first two non-Gaussian shape descriptors.
\begin{theorem}[Why first--fourth moments suffice, and why kurtosis matters]
\label{thm:why_first_fourth}
Let $X$ denote a high-frequency DCT coefficient drawn from the residual-frequency representation used by our detector,
and let $P_r$ and $P_f$ be the distributions of $X$ for real and adversarially generated images, respectively. Define standardized moments
\[
\mu_k \;=\; \mathbb{E}\!\left[\left(\frac{X-\mu}{\sigma}\right)^k\right],
\quad \mu=\mathbb{E}[X],\ \sigma^2=\mathrm{Var}(X)>0.
\]
Assume: (i) $P_r$ is approximately Gaussian after standardization (sensor-noise-dominated residuals), and
(ii) $P_f$ differs from $P_r$ primarily by sparse, high-magnitude spectral outliers (a ``spike'' deviation), i.e.,
a stylized model $Z\sim(1-\varepsilon)\mathcal{N}(0,1)+\varepsilon\mathcal{N}(0,\tau^2)$ with $\varepsilon\ll 1$ and $\tau^2\gg 1$
captures the dominant departure in the tail of standardized coefficients.

\noindent Then:

\noindent\textbf{(a) First--fourth moments are the right set.}
Under (i), $P_r$ is fully characterized (up to estimation error) by $\mu_1=0$ and $\mu_2=1$ and has $\mu_3=0,\ \mu_4=3$,
so the first four moments capture all low-order non-Gaussian departures from $P_r$ via $\mu_3$ (asymmetry) and $\mu_4$ (tail-heaviness).

\noindent\textbf{(b) Kurtosis is especially important.}
Under (ii), the excess kurtosis satisfies
\[
\kappa \;:=\; \mu_4-3 \;=\; \Theta(\varepsilon\,\tau^4),
\quad \text{whereas}\quad
\mathrm{Var}(Z) \;=\; 1+\Theta(\varepsilon\,\tau^2).
\]
Hence, for rare but large outliers, $\mu_4$ grows parametrically faster than any first- or second-order statistic,
making kurtosis the \emph{lowest-order} moment that sharply exposes spike-induced tail deviations.

\noindent\textbf{(c) Higher moments are unnecessary (and often undesirable) at patch scale.}
For $n$ coefficient samples (e.g., from a fixed DCT block/band), the empirical moment estimator
$\widehat{\mu}_k=\frac{1}{n}\sum_{i=1}^n Z_i^k$ satisfies
\[
\mathrm{Var}(\widehat{\mu}_k)
\;=\;\frac{1}{n}\Big(\mathbb{E}[Z^{2k}]-(\mathbb{E}[Z^k])^2\Big),
\]
so reliable estimation of $\mu_k$ requires control of the $2k$-th moment.
In heavy-tailed settings, $\mathbb{E}[Z^{2k}]$ grows rapidly with $k$,
implying that moments beyond fourth order ($k\ge 5$) become statistically unstable at practical $n$,
while $\mu_4$ remains the highest-order moment that is both tail-sensitive and reliably estimable.
\end{theorem}
\paragraph{Proof sketch.}
Part (a) follows from Gaussian standardized-moment identities and the fact that $\mu_3,\mu_4$ are the first moments
that quantify non-Gaussian asymmetry and tail behavior.
Part (b) follows by direct computation of $\mathbb{E}[Z^2]$ and $\mathbb{E}[Z^4]$ under the spike mixture,
showing $\kappa$ scales as $\varepsilon\tau^4$ while variance scales as $\varepsilon\tau^2$.
Part (c) follows from $\mathrm{Var}(\widehat{\mu}_k)=\mathrm{Var}(Z^k)/n$ and the rapid growth of $\mathbb{E}[Z^{2k}]$ for heavy tails,
which makes high-order moment estimation noisy under fixed sample budgets.

\begin{figure}[!htb]
    \centering

    \begin{subfigure}{0.45\linewidth}
        \centering
        \includegraphics[width=\linewidth]{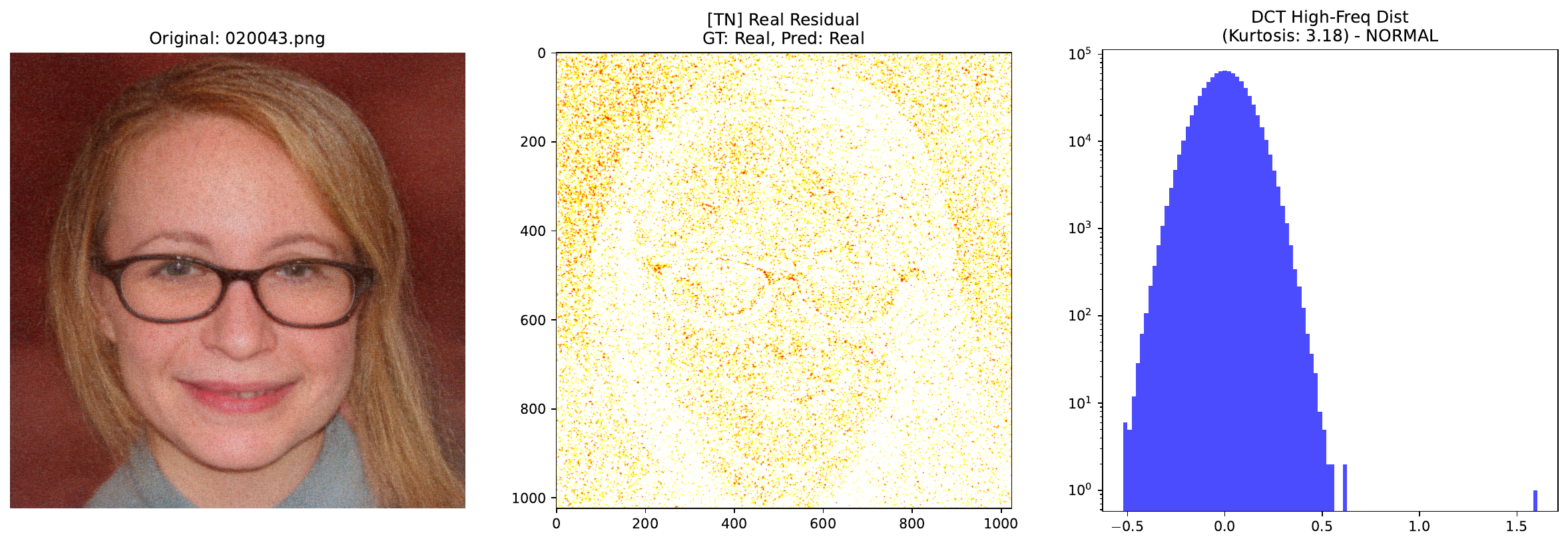}\par
        \includegraphics[width=\linewidth]{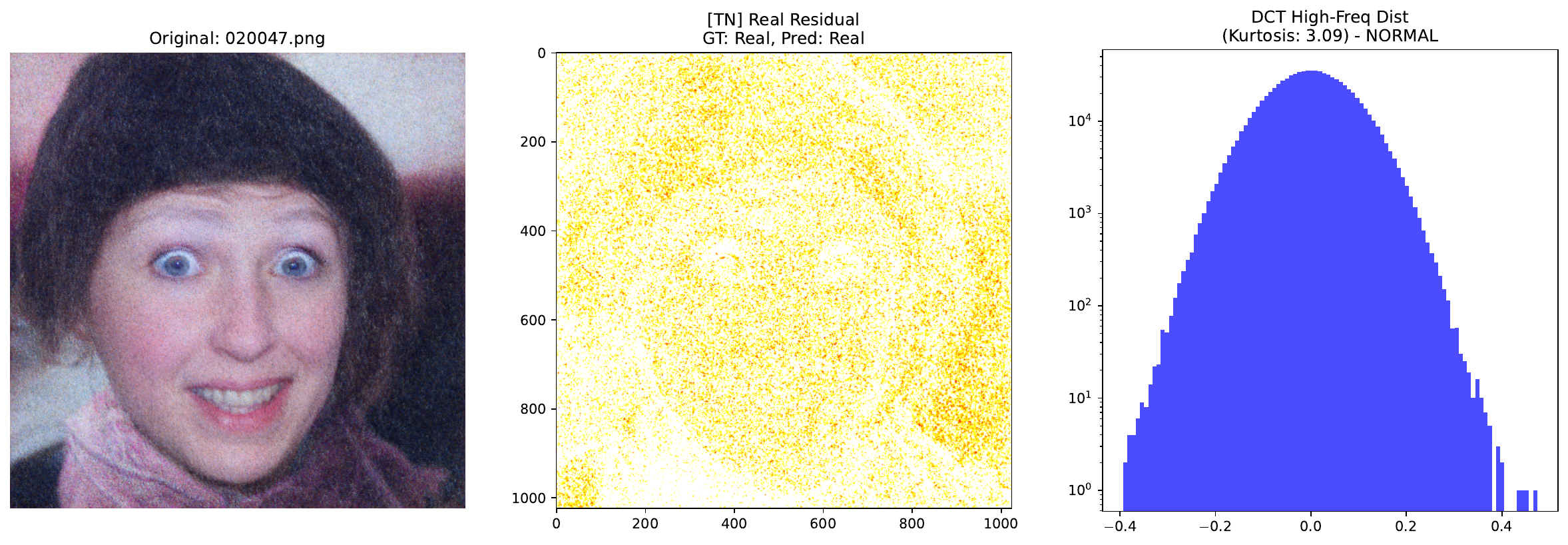}\par
        \includegraphics[width=\linewidth]{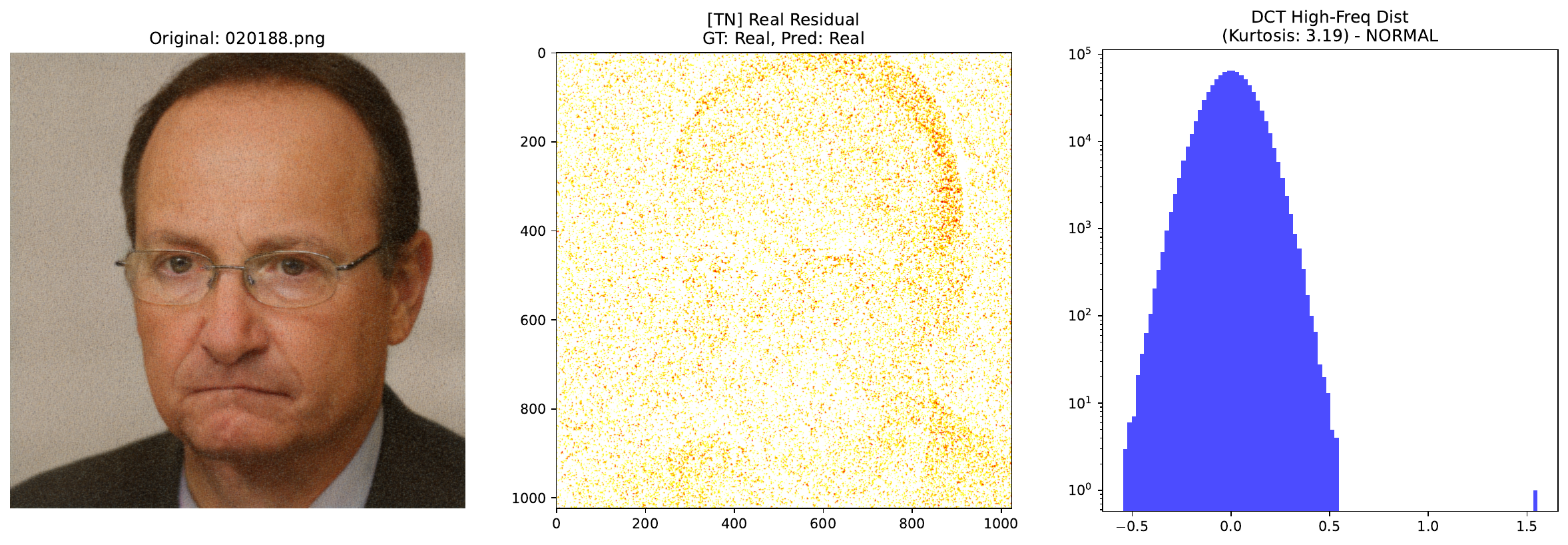}\par
        \includegraphics[width=\linewidth]{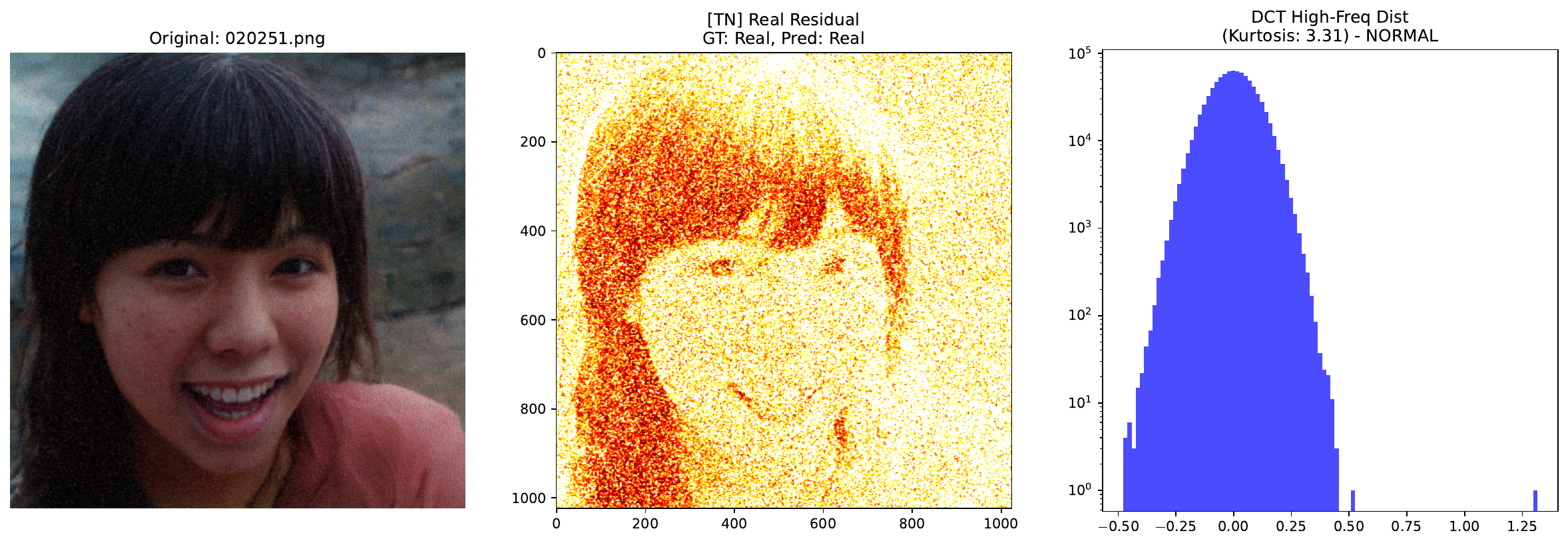}\par
        \includegraphics[width=\linewidth]{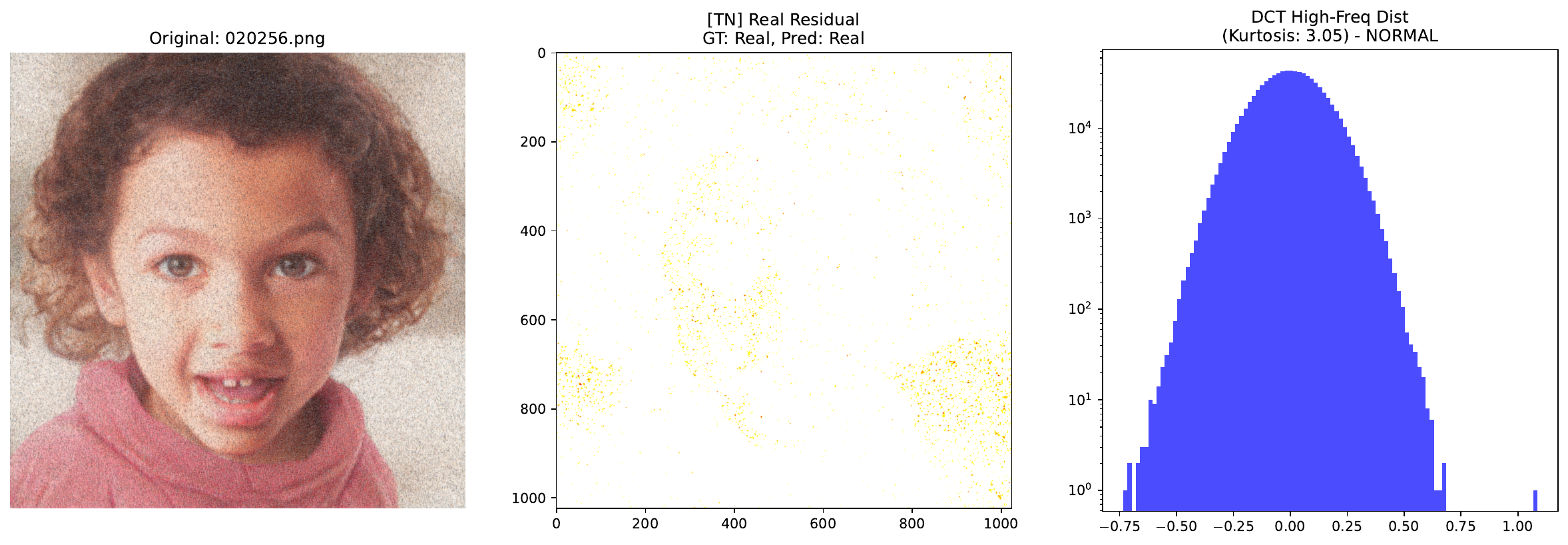}\par
        \includegraphics[width=\linewidth]{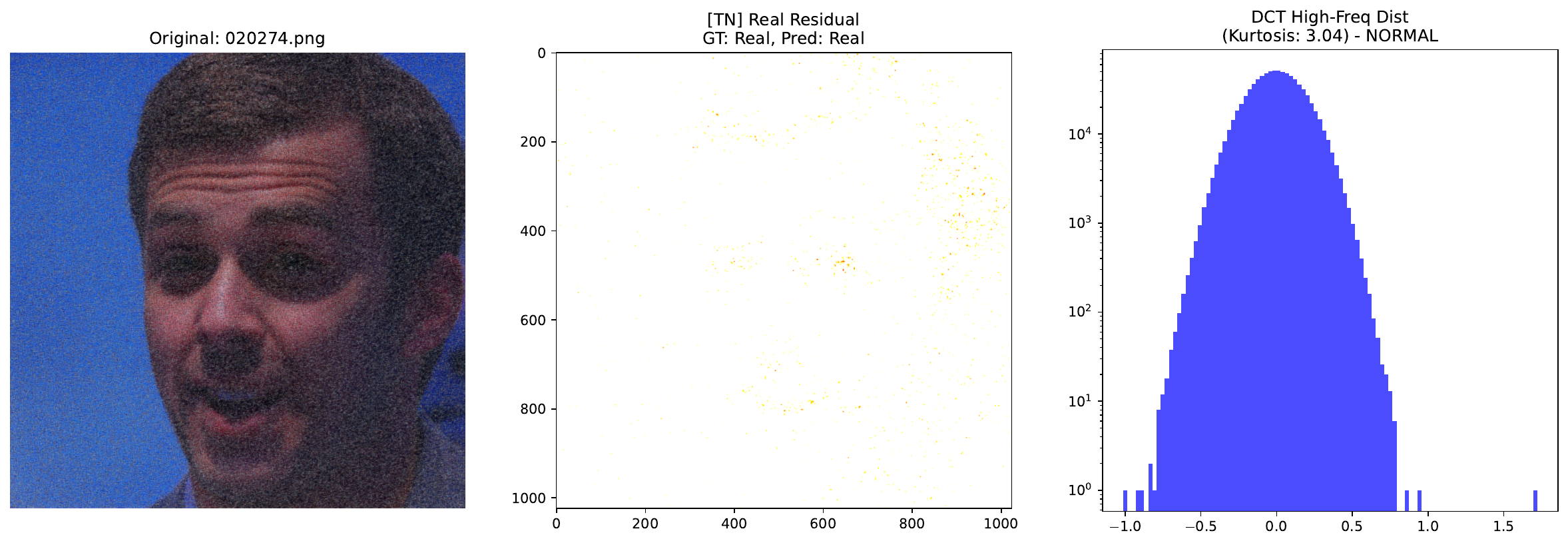}\par
        \includegraphics[width=\linewidth]{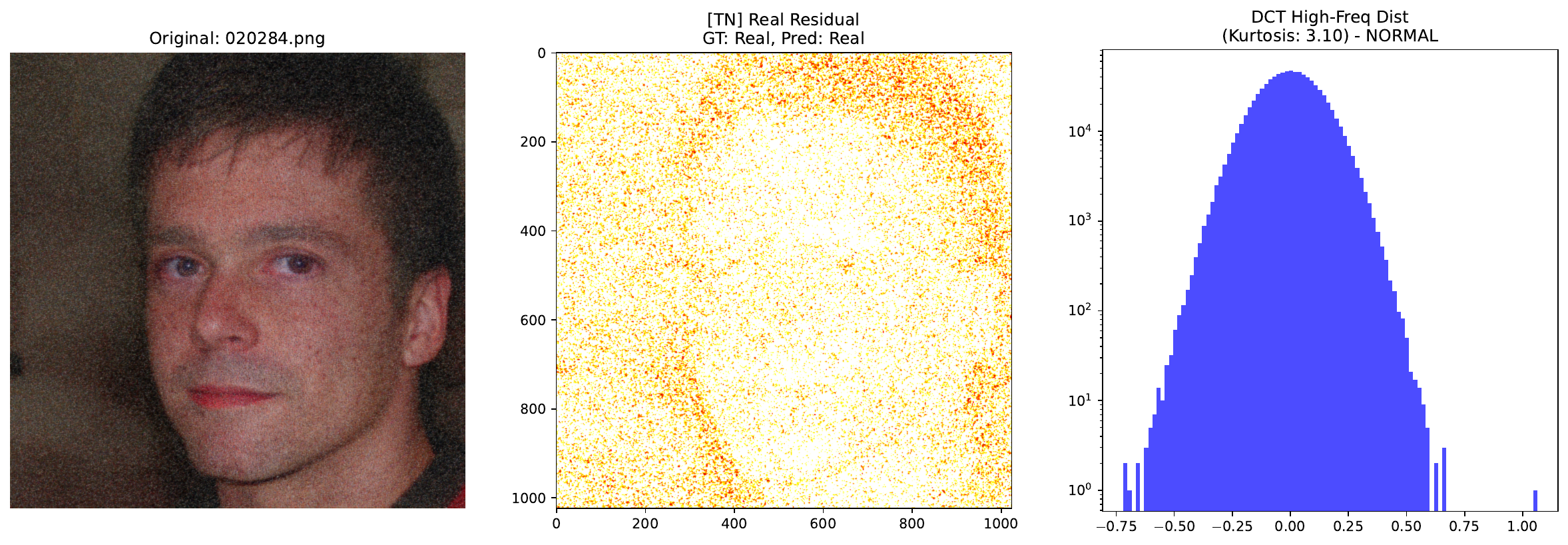}
        \caption{Real image residuals}
        \label{fig:res_real}
    \end{subfigure}
    \hfill
    \begin{subfigure}{0.45\linewidth}
        \centering
        \includegraphics[width=\linewidth]{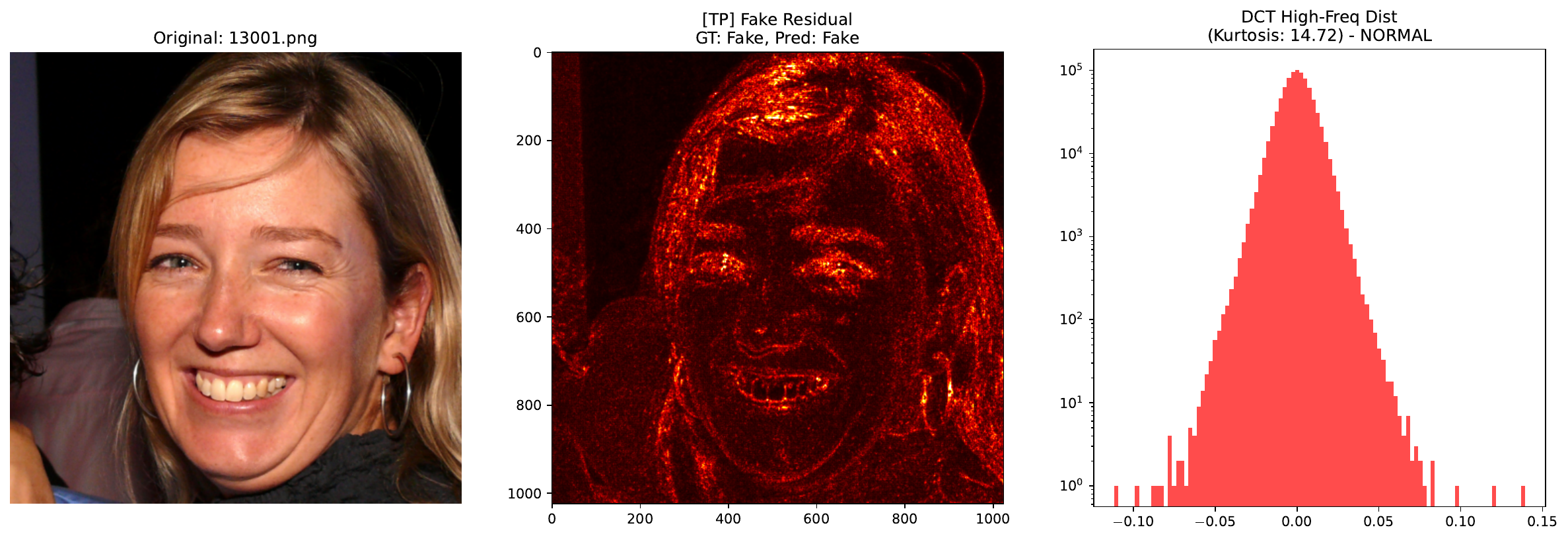}\par
        \includegraphics[width=\linewidth]{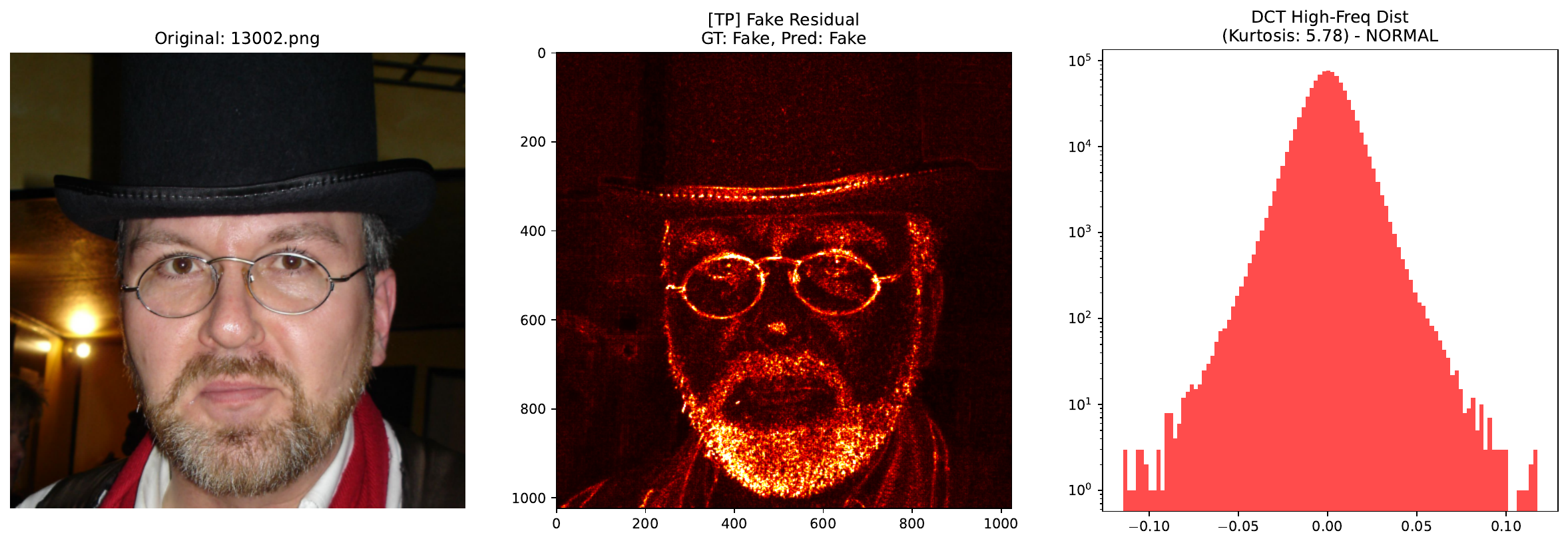}\par
        \includegraphics[width=\linewidth]{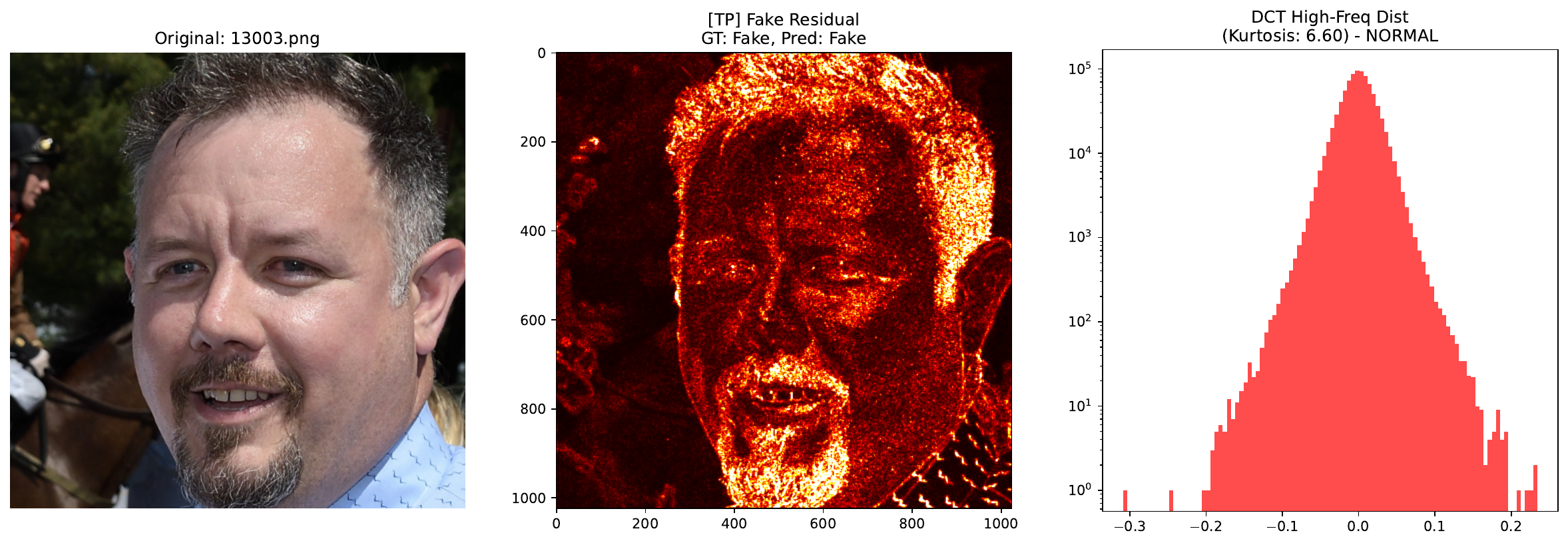}\par
        \includegraphics[width=\linewidth]{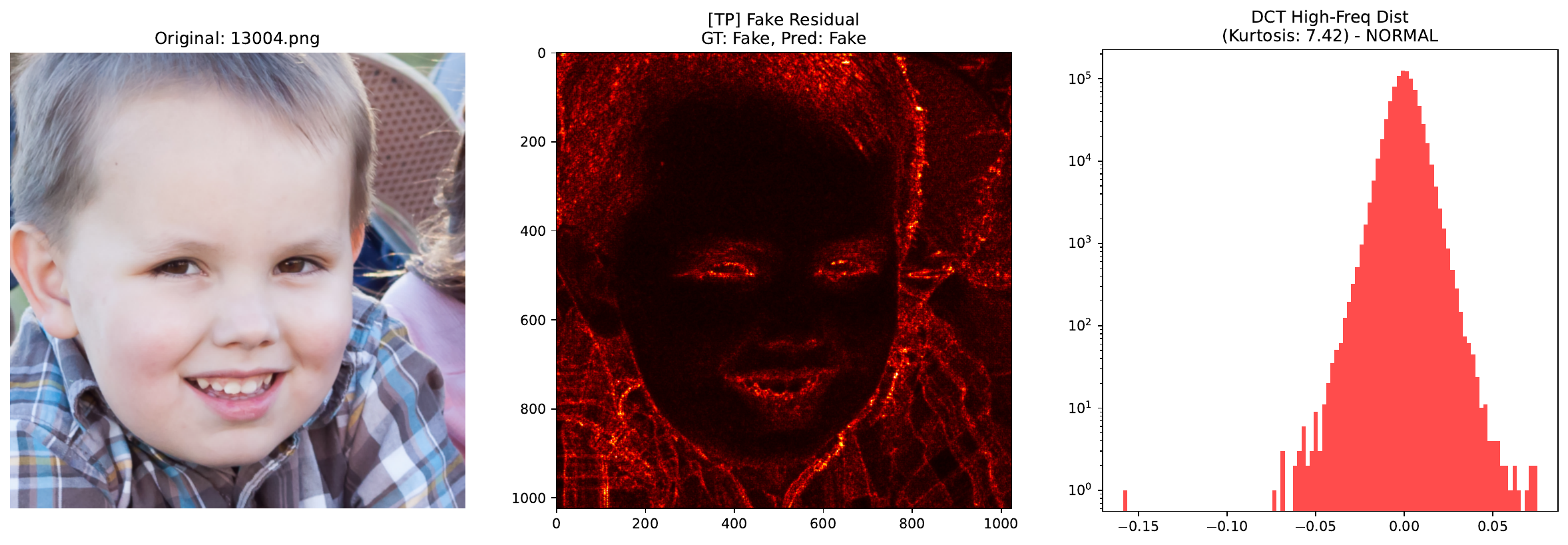}\par
        \includegraphics[width=\linewidth]{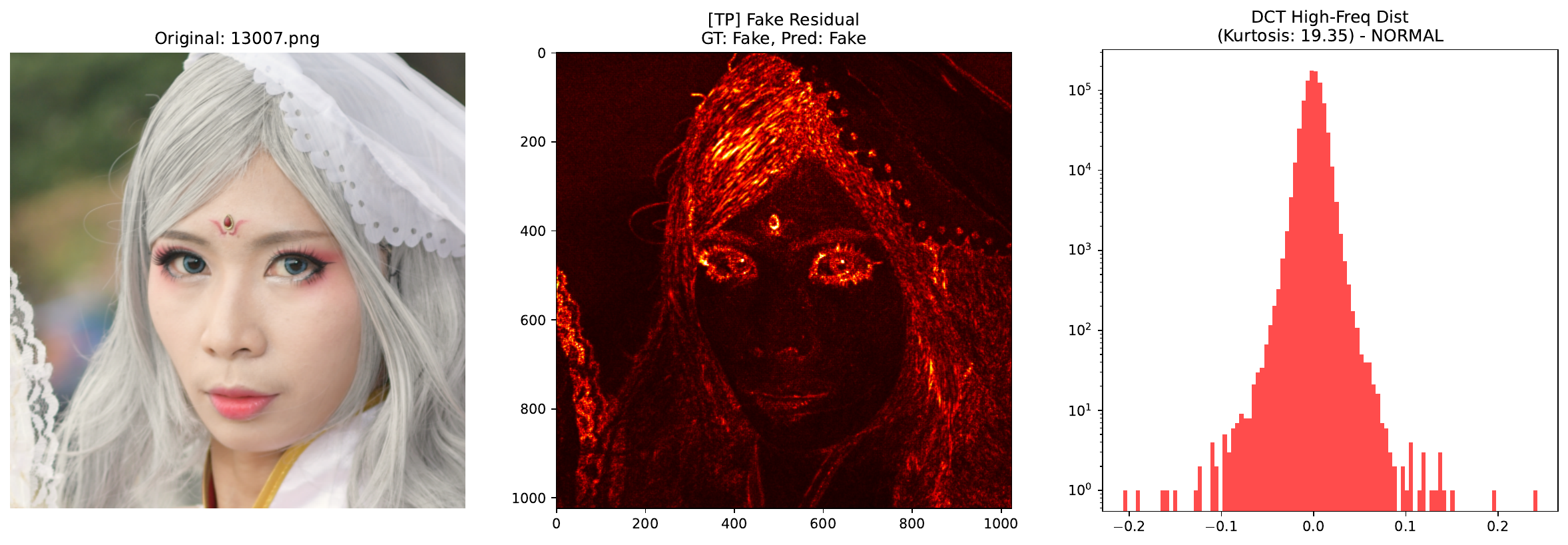}\par
        \includegraphics[width=\linewidth]{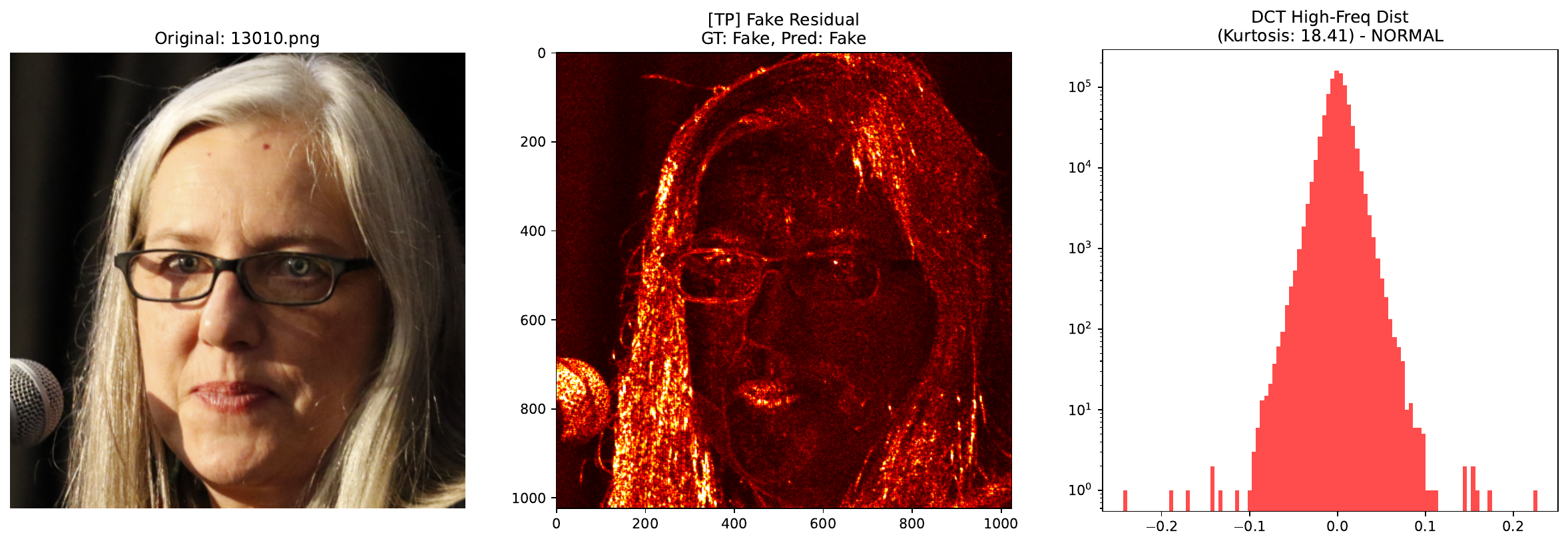}\par
        \includegraphics[width=\linewidth]{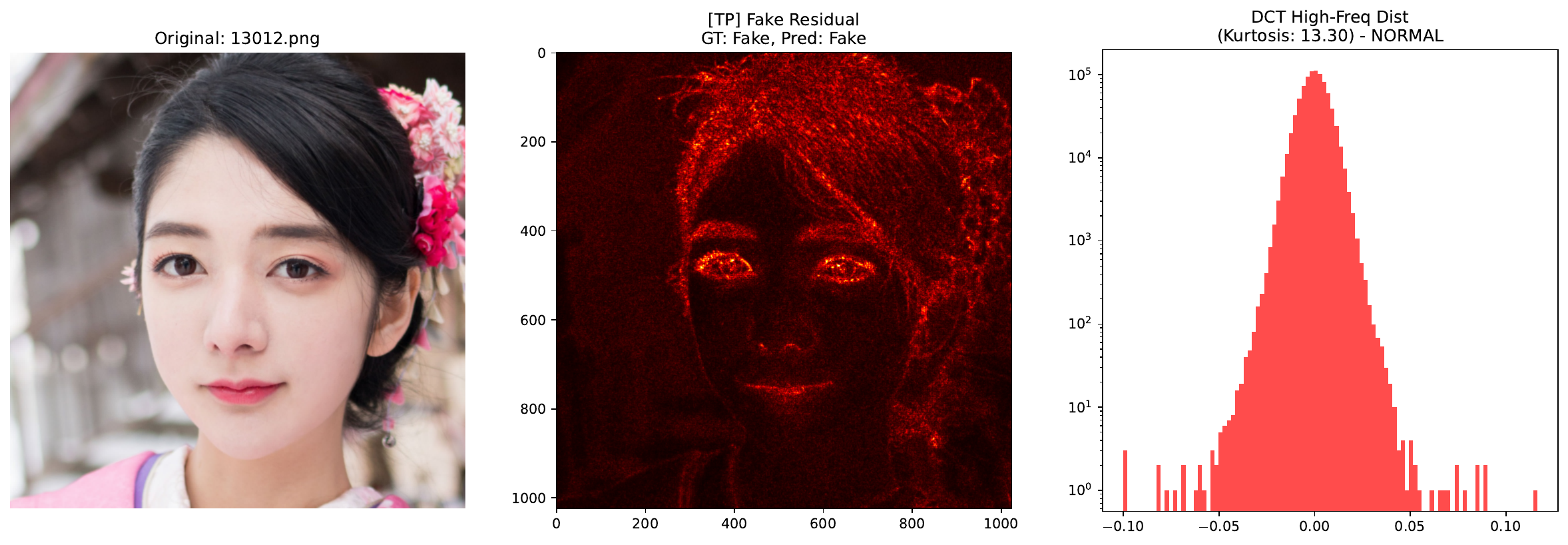}
        \caption{Adversarial fake image residuals}
        \label{fig:res_fake}
    \end{subfigure}

    \caption{Comparison of spatial and frequency characteristics of residuals across multiple samples. 
(a) Residuals from real images exhibit near-Gaussian behavior, consistent with low-amplitude, stochastic noise. 
(b) Residuals from adversarially generated images display consistently heavy-tailed (leptokurtic) distributions, indicative of structured high-frequency artifacts.}
    \label{fig:main_comparison}
\end{figure}

\begin{figure*}[!htb]
\centering
\includegraphics[width=1.3in,angle=90]{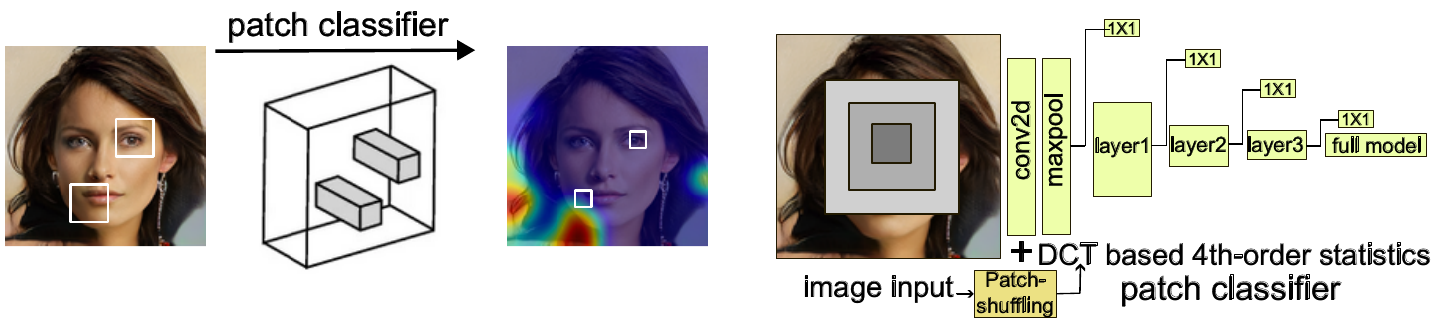}
\caption{Modified Patch-Forensics architecture}
\label{Fig:network_architecture_patch-forensics}
\end{figure*}

\section{Analysis of Higher-Order Statistical Moments and Other Architectural Figures}
\label{sec:kurtosis-avg}
Figure~\ref{fig:res_real} presents the spatial residuals obtained from seven true negative (real) samples. Across multiple instances, these residuals exhibit low-amplitude, spatially unstructured behavior that is characteristic of stochastic noise. Consistent with this qualitative observation, the corresponding high-frequency DCT coefficient distributions yield kurtosis values of $3.18$, $3.09$, $3.19$, $3.31$, $3.05$, $3.04$, and $3.10$, all of which closely approximate the kurtosis of a normal distribution ($\kappa \approx 3$). The average kurtosis across all 741 true negative samples (detected by our modified Resynthesis) is $3.17$, indicating that the reconstruction errors in real images are dominated by random sensor noise and lack systematic high-frequency structure.

In contrast, the seven adversarial fake samples shown in Figure~\ref{fig:res_fake} exhibit markedly different higher-order statistical behavior. Their high-frequency DCT residuals display substantially elevated kurtosis values of $14.72$, $5.78$, $6.60$, $7.42$, $19.35$, $18.41$, and $13.30$, reflecting strongly leptokurtic, heavy-tailed distributions. This pronounced deviation from Gaussianity is indicative of structured and sporadic high-frequency artifacts. The average kurtosis across all 925 true positive/fake samples (detected by our modified Resynthesis) reaches $22.4$, underscoring a clear statistical separation from real images. Such heavy-tailed residual distributions are consistent with known artifacts introduced by generative synthesis pipelines. While these adversarial images may appear visually indistinguishable from real images to human observers, the elevated higher-order moments in the frequency domain reveal persistent generative traces. Crucially, these signatures are reliably captured by our proposed model, enabling robust discrimination between authentic and adversarial content beyond what is possible through first- and second-order statistics alone.


\begin{figure}[!htb]
\centering
\includegraphics[width=3.73in,angle=90]{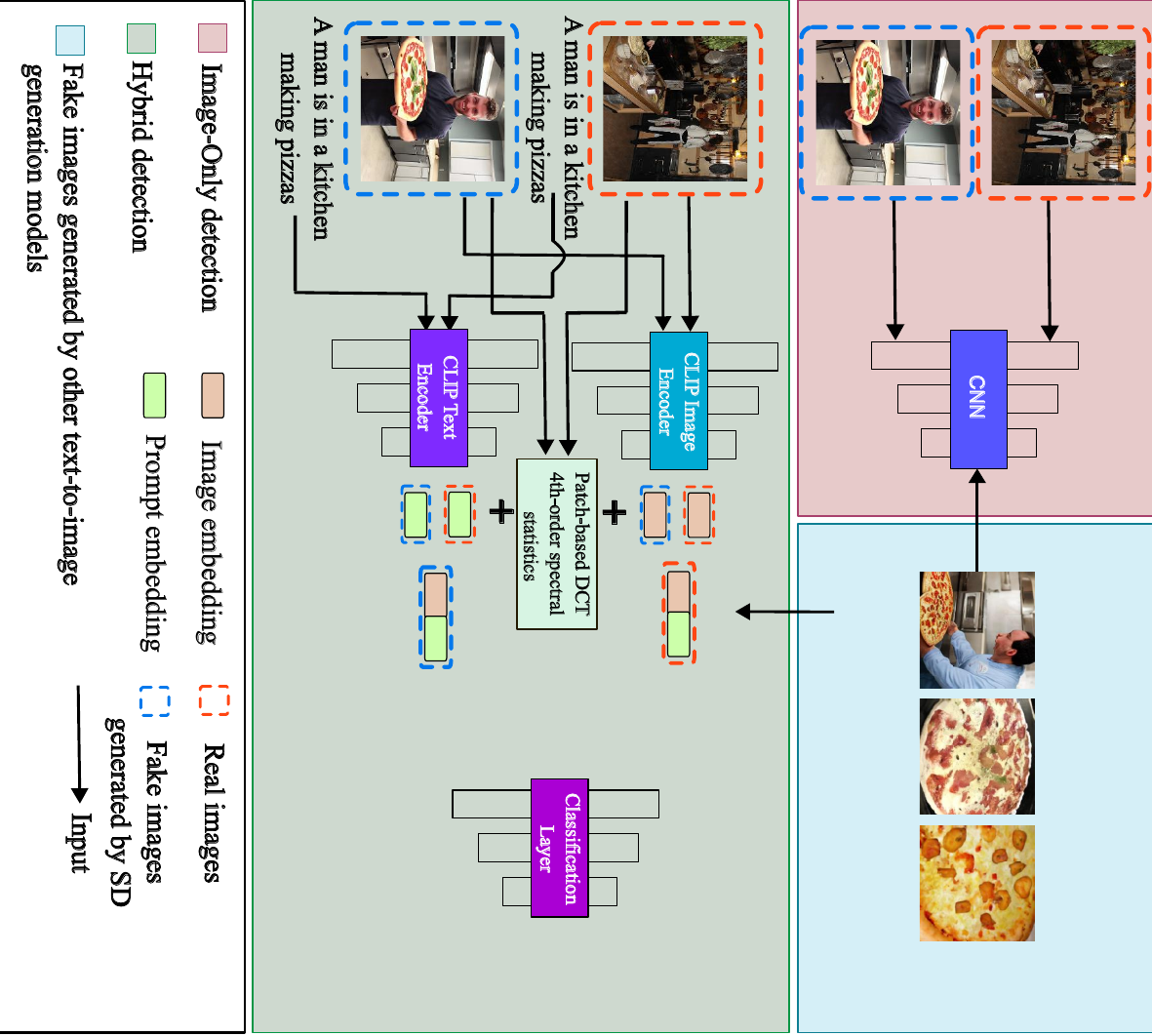}
\caption{Modified DE-FAKE architecture}
\label{Fig:network_architecture_de-fake}
\end{figure}

\begin{figure}[!htb]
\centering
\includegraphics[width=1.7in,angle=90]{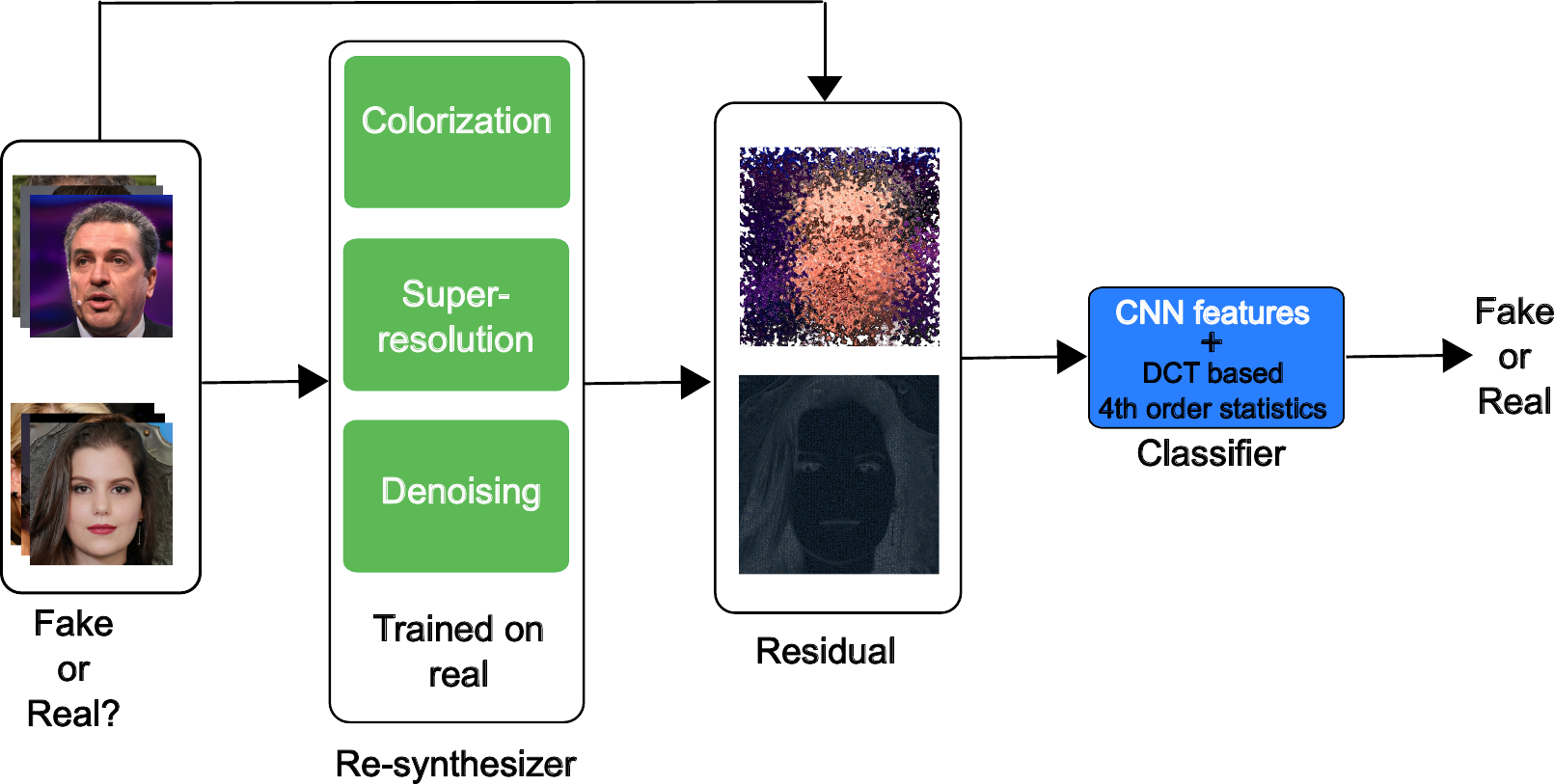}
\caption{Modified Resynthesis architecture}
\label{Fig:network_architecture_Re-synthesis}
\end{figure}

\end{document}